\documentclass[a4paper,10pt,romanappendices
]{IEEEtran}

\usepackage{commath,amsmath,graphicx,epstopdf,amsthm,amssymb,float,cite,multicol,enumitem,color}

\newcommand{\avec}{{\bf{a}}}

\newcommand{\xvec}{{\bf{x}}}

\newcommand{\rvec}{{\bf{r}}}

\newcommand{\svec}{{\bf{s}}}
\newcommand{\vvec}{{\bf{v}}}

\newcommand{\onevec}{{\bf{1}}}
\newcommand{\zerovec}{{\bf{0}}}

\newcommand{\Amat}{{\bf{A}}}
\newcommand{\Bmat}{{\bf{B}}}
\newcommand{\Cmat}{{\bf{C}}}

\newcommand{\Imat}{{\bf{I}}}

\newcommand{\Pmat}{{\bf{P}}}

\newcommand{\Xmat}{{\bf{X}}}

\newcommand{\Diag}{{\rm Diag}}

\def\btheta{$\boldsymbol{\theta}$}

\def\btheta{{\mbox{\boldmath $\theta$}}}

\def\bvarphi{{\mbox{\boldmath $\varphi$}}}

\def\thetavec{{\boldsymbol{\theta}}}

\newcommand{\be}{\begin{equation}}
\newcommand{\ee}{\end{equation}}
\newcommand{\beqna}{\begin{eqnarray}}
\newcommand{\eeqna}{\end{eqnarray}}

\makeatletter
\def\user@resume{resume}
\def\user@intermezzo{intermezzo}
\newcounter{previousequation}
\newcounter{lastsubequation}
\newcounter{savedparentequation}
\makeatother 
\newcommand{\diff}{{\textnormal{d}}}

\newcommand{\sm}{\hat{\sigma}_m^2}
\newcommand{\Rxm}{\hat{\textbf{R}}_{\textbf{x},m}}
\newcommand{\RxM}{\hat{\textbf{R}}_{\textbf{x},M}}
\newcommand{\psimML}{\hat{\psi}^{(m)}_{ML}}
\newcommand{\Rxminv}{\hat{\textbf{R}}_{\textbf{x},m}^{-1}}

\newcommand{\thm}{\boldsymbol{\theta}^{(m)}}
\newcommand{\thmtr}{\boldsymbol{\theta}^{(m)}_{tr}}

\newcommand{\thML}{\hat{\boldsymbol{\theta}}^{(m)}_{ML}}

\newcommand{\thetaYW}{\hat{\boldsymbol{\theta}}^{(m)}_{YW}}
\newcommand{\thmz}{\boldsymbol{\theta}^{(m)}_0}

\DeclareMathOperator*{\minm}{\min}

\DeclareMathAlphabet\mathbfcal{OMS}{cmsy}{b}{n}

\title{Model Selection via MCRB Optimization}
\author{
Nadav E. Rosenthal, \emph{Student Member, IEEE}, and Joseph Tabrikian,~\IEEEmembership{Fellow,~IEEE}
 \vspace{-20pt}
\thanks{{This research was partially supported by THE ISRAEL SCIENCE FOUNDATION (grant No. 2666/19).\newline
\indent The authors are with the School of Electrical and Computer Engineering,
Ben-Gurion University of the Negev, Beer-Sheva 84105, Israel (e-mail: rosenthn@post.bgu.ac.il; joseph@bgu.ac.il).
}
}

}
\begin{document}

\maketitle
\nopagebreak

\begin{abstract} 
In many estimation theory and statistical analysis problems, the true data model is unknown, or partially unknown. To describe the model generating the data, parameterized models of some degree are used. A question that arises is which model should be used to best approximate the true model, a.k.a. model selection. In the field of machine learning, it is encountered in the form of architecture types of neural networks, number of model parameters, etc. In this paper, we propose a new model selection criterion, based on the misspecified Cram\'{e}r-Rao bound (MCRB) for mean-squared-error (MSE) performance. The criterion selects the model in which the bound on the estimated parameters MSE is the lowest, compared to other candidate models. Its goal is to minimize the MSE with-respect-to (w.r.t.) the model. The criterion is applied to the problems of direction-of-arrival (DOA) estimation under unknown clutter / interference, and spectrum estimation of auto-regressive (AR) model. It is shown to incorporate the bias-variance trade-off, and outperform the Akaike information criterion (AIC), finite-sample corrected AIC (AICc), and minimum description length (MDL) in terms of MSE performance.
\end{abstract}

\section{Introduction}
\label{sec:NBE:intro}
In many statistical signal processing applications, the true data model is unknown, and a parameterized model is used to represent the data. In some other applications, the true data model may be known, but the corresponding estimators, which perfectly rely on the complete model, may be complex or may result on poor performance due to overfitting. Accordingly, one of the main challenges in statistical signal processing is model selection, which deals with finding a parameterized model which reliably represent the data. Some well-investigated examples are linear predictive coding (LPC) in speech and audio processing, polynomial curve fitting, 
and parametric and non-parametric spectrum estimation. 
In the field of machine learning, model selection requires choosing an architecture that describes the relation between an input sample and system output (sometimes called a label), and optimizing the system by tuning the model parameters.  

Estimation accuracy in signal processing and machine learning algorithms depends on several factors, such as, how well the chosen model describes the data, the number of model parameters, and the number of available samples or training data used to estimate the model parameters.
On the one hand, under a given training data size, high-order models with large number of model parameters may suffer from the overfitting effect. On the other hand, simple low-order models can result in model misspecification. Both cases may lead to significant estimation errors. Therefore, model selection techniques are required. 

Model order selection techniques are surveyed  
in \cite{Ding, Stoica, Kadane}, and include the Akaike information criterion (AIC)\cite{AIC}, finite-sample corrected AIC (AICc), Bayesian information criterion (BIC) \cite{BIC}, and minimum description length (MDL) \cite{MDL1, MDL2, MDL3, MDL4}. AIC is asymptotically efficient in the non-parametric framework, meaning that when the number of samples tends to infinity, the selected model's performance is equivalent to the best offered by the candidate models. BIC is consistent in the parametric framework, in the manner that it selects the smallest order true model from the candidate models. Though, AIC is inconsistent in the parametric framework, as it tends to overfit when more than one candidate models can describe the true model. Moreover, BIC is not asymptotically efficient in a non parametric framework. Both criteria are based on the likelihood function, as they try to select the best model is some manner, but they are not optimal in terms of mean-squared-error (MSE) performance.

We propose to use performance bounds on the estimation MSE as a criterion for model selection. In most signal processing problems, the impact of underfitting is more pronounced than that of overfitting.  In the case of overfitting, the assumed model reliably describes the true model, but it includes additional parameters which need to be estimated. Therefore, the expected performance can be evaluated via conventional MSE bounds. However, in the case of underfitting, the assumed model is incorrect, and therefore, conventional MSE bounds which do not consider model misspecification, cannot provide reliable performance prediction. Accordingly, we propose to using misspecified MSE bounds, as a criterion for model selection. 
The model selection approach, firstly described in \cite{Rosenthal}, would be to minimize the MSE bound for parameter estimation with respect to (w.r.t.) the model. 

In the recent years, model misspecification and its limitations on estimation performance analysis has been intensively investigated in the signal processing community \cite{Stefano3}. The widely used Cram\'{e}r–Rao bound (CRB)\cite{CRB1,CRB2} and its applications in system design and analysis, encouraged researchers to derive non-Bayesian and Bayesian bounds under model misspecification. In \cite{White,White_Book} the asymptotic properties of the maximum-likelihood estimator (MLE) were studied under misspecified models where the data samples are statistically independent (see also \cite{Noam_Tabrikian}).
In \cite{Vuong86} a Cram\'{e}r-Rao type bound for parameter estimation was derived under misspecified models, introducing the concept of misspecified unbiasedness (MS-unbiased). The works in \cite{Richmond,Richmond2} studied the effects of modeling misspecification on the MSE of the MLE (see also \cite{White,White_Book}) and derived the misspecified CRB (MCRB), under other constraints. The MCRB extends the CRB to misspecified scenarios, depicting its impact on the estimation bias and covariance. The MCRB was applied to various signal processing problems in \cite{Stefano,Stefano3,Wang_Gini,9537597,MCPHEE2023108872,https://doi.org/10.48550/arxiv.2203.03398}, and extended to constrained problems in \cite{Stefano2, Fortunati}.

In this paper, we propose a model selection criterion, focused on minimizing the MSE of estimated parameters in the model. The candidate models are presented in two forms. In the first type, nuisance parameters are estimated using training data, and the parameters of interest are estimated using test data. In the second type, the parameters of interest are mapped by the estimated parameters. The model selection criterion evaluates the MCRB for the candidate models, and selects the model with the lowest MCRB, which is the optimal in terms of MSE performance. The MCRB incorporates the bias-variance trade-off, as underfitted models result in large bias errors and low variance, while overfitted models result in small bias errors and large variance.
The approach is applied to direction-of-arrival (DOA) estimation under unknown clutter / interference,
and parametric spectrum estimation of auto-regressive (AR) model. It is shown to outperform in terms of MSE performance other model selection criteria. 
The main contributions of this paper are:
\begin{itemize}
    \item Derivation of the proposed model selection criterion, based on the MCRB for MSE performance.
    \item Application to DOA estimation under unknown clutter / interference, including MCRB derivation and study.
     \item Application to spectrum estimation of AR model, including MCRB derivation and study.
\end{itemize}
This paper is organized as follows. Section \ref{Formulation} presents the problem and the MSE criterion as a function of the model. In Section \ref{sec:MCRB for model order selection}, the proposed model selection criterion is derived. Section \ref{Applications} presents model selection applications, and Section \ref{Simulations} presents simulation results.
Finally, our conclusions are summarized in Section \ref{Conclusion}.

Boldface lowercase and boldface uppercase letters are used to denote vectors and matrices, respectively. Unbold letters of either lowercase or uppercase are used for scalars. Superscripts $^T$, $^H$, and $^*$ stand for transpose, conjugate transpose, and conjugation operations, respectively. The real and imaginary operators will be denoted by $\Re[\cdot],\Im[\cdot]$, respectively. The gradient of a scalar $b$ w.r.t. $\mathbf{a}\in\mathbb{R}^K$ is a column vector, whose $j$th element is defined as $\left[\frac{\diff b(\mathbf{a})}{\diff \mathbf{a}}\right]_{j}\triangleq\frac{\partial b(\mathbf{c})}{\partial c_j}\Big|_{\mathbf{c}=\mathbf{a}}$. 
The Hessian matrix of a scalar $b$ w.r.t. $\mathbf{a}$ is a $K\times K$ matrix defined by $\frac{\diff^2 b(\mathbf{a})}{\diff\mathbf{a}\diff^T\mathbf{a}}\triangleq\frac{\diff}{\diff\mathbf{a}}\left(\frac{\diff b}{\diff^T\mathbf{a}}\right)$. 
Given a vector $\mathbf{b}$, its derivative w.r.t. $\mathbf{a}$ is a matrix whose $j,k$-th entry is defined as $\left[\frac{\diff \mathbf{b}}{\diff \mathbf{a}}\right]_{j,k}\triangleq\frac{\partial b_j(\mathbf{c})}{\partial c_k}\Big|_{\mathbf{c}=\mathbf{a}}$.  
$\Diag(\avec)$ is a diagonal matrix whose diagonal elements are the entries of the vector $\avec$, and $tr(\Amat), det(\Amat)$ are the trace and determinant of the matrix $\Amat$, respectively. Column vectors of size $N$, whose entries are equal to 0 or 1 are denoted by $\zerovec_N$ and $\onevec_N$, respectively, and identity matrix of size $N\times N$ is denoted by $\Imat_N$. 

\section{Problem Formulation}
\label{Formulation}
\indent In this section, we first provide the basic notations and definitions, used throughout the paper. In Subsection \ref{MSE_criterion} the MSE criterion is presented, as a function of the model.

\subsection{Problem Preliminaries}\label{ssec:NBE:pre}
Let $\xvec \in \Omega_{\xvec} \subseteq \mathbb{R}^{J}$ be a random vector representing the observation vector, and $g(\xvec;\bvarphi)$ denote the true PDF of $\xvec$, which belongs to a family of distributions $\mathcal{G}=\{g\left(\xvec;{\bvarphi}\right)| \bvarphi \in \Omega_{\bvarphi} \subseteq \mathbb{R}^s\}$ . $\bvarphi$ is an unknown deterministic vector, which is the parameter of interest. In cases where the true PDF $g(\xvec;\bvarphi)$ is not known, it can be approximated with a parameterized PDF known up to a set of parameters to be determined in a learning stage. Two types for parameterized PDFs are described, which are used later.
\begin{itemize}
    \item Type 1: $\mathcal{O}^{(m)} = \{ f^{(m)}(\xvec;\bvarphi,\thm) | \bvarphi \in \Omega_{\bvarphi}, \thm \in \Omega^{1}_{\thetavec^{(m)}} \}$
    \item Type 2: $\mathcal{O}^{(m)} = \{ f^{(m)}(\xvec;\thm) | \thm \in \Omega^{2}_{\thetavec^{(m)}} \}$
\end{itemize}
Let $\mathcal{O}^{(m)}$ be the $m$th model, where $m \in \mathcal{M} $. This model is defined by the set of parametric PDFs $f^{(m)}(\xvec;\bvarphi,\thm)$ for type 1 or $f^{(m)}(\xvec;\thm)$ for type 2, with parameter spaces $\Omega^1_{\thetavec^{(m)}}, \Omega^2_{\thetavec^{(m)}}$, respectively. In the former, the model parameters $\thm$ are pre-determined using a training set $\{\xvec_t, \bvarphi_t \}$, and afterwards $\bvarphi$ can be estimated using another measurement $\xvec$, while in the latter $\thm$ are estimated using $\xvec$. For type 2, the parameter of interest is defined by $\bvarphi^{(m)}\left(\thm\right),$ where $\bvarphi^{(m)} : \Omega^2_{\thetavec^{(m)}} \rightarrow \mathbb{R}^s$ is a continuously differentiable mapping. Generally, $m$ denotes the model index, which is not necessarily the model order. A model class is defined by the collection of models $\{\mathcal{O}^{(m)}\}_{m \in \mathcal{M}}$. Note that $g(\xvec;\bvarphi)$ is not necessarily included in the model class, that is, there may not exist any $m^* \in \mathcal{M}$ and $\boldsymbol{\theta}^{(m*)} \in \Omega^1_{\thetavec^{(m*)}}$ such that $g(\xvec;\bvarphi)=f^{(m*)}(\xvec;\bvarphi,\boldsymbol{\theta}^{(m*)})$ or $\boldsymbol{\theta}^{(m*)} \in \Omega^2_{\thetavec^{(m*)}}$ such that $g(\xvec;\bvarphi)=f^{(m*)}(\xvec;\boldsymbol{\theta}^{(m*)})$.

Type 1 describes parametric models in which nuisance parameters $\thm$ are included, while separated from the parameters of interest $\bvarphi$.
This problem statement can be used to formulate a large class of learning systems, where the estimate of $\bvarphi$ represents the output of the system and $\thm$ stands for the machine or network parameters. Type 2 describes parametric models in which the parameters of interest $\bvarphi$ can be mapped by  $\bvarphi^{(m)}(\cdot)$ and the estimated parameters $\thm$, like linear models.

\subsection{MSE Criterion}
\label{MSE_criterion}
For deriving an estimator of $\bvarphi$, we approximate the data model by $f^{(m)}(\xvec;\bvarphi,\thm)$ for type 1 or $f^{(m)}(\xvec;\thm)$ for type 2.
In the former, an estimator of 
$\bvarphi$ with index $m$ by $\xvec$ is $\hat{\bvarphi}^{(m)}\left (\xvec,\thmtr \right)$, where $\thmtr$ are pre-determined using a training set $\{\xvec_t,\bvarphi_t\}_{t=1}^T$.
In the latter, an estimator of 
$\bvarphi$ with order $m$ by $\xvec$ is
$\bvarphi^{(m)}\left(\hat{\boldsymbol{\theta}}^{(m)}(\xvec)\right)$. For simplicity, the estimators in both types are referred by $\hat{\bvarphi}^{(m)}$.
The estimation error of $\bvarphi$ under the $m$th model is defined by:
\begin{equation}\label{error}
    \textbf{e}(m) = \bvarphi - \hat{\bvarphi}^{(m)},
\end{equation}
and the corresponding MSE matrix of $\bvarphi$ with model order $m$ is:
\begin{equation}\label{MSE phi tau}\begin{split}
    \textbf{MSE}(m) &= {\rm E } _g \left [\textbf{e}(m)
    \textbf{e}^T(m) \right ], 
\end{split}
\end{equation}
where ${\rm E } _g [\cdot]$ represents expectation with PDF $g(\xvec;\bvarphi)$.
In order to determine the optimal model, we will derive a lower bound on the MSE using the MCRB under each model and minimize it with respect to (w.r.t.) the model. 
\section{Model Selection via MCRB Minimization}\label{sec:MCRB for model order selection}
In this section, we derive a model selection criterion using the MCRB.
For the derivation of the MCRB, we assume that the regularity conditions used in \cite{Huber,Vuong86,Stefano,Richmond} are satisfied $\forall m \in \mathcal{M}$, i.e. for all the models in the class $\{\mathcal{O}^{(m)}\}_{m \in \mathcal{M}}$.
Thus, the pseudo-true parameter vector which minimizes the Kullback–Leibler divergence (KLD) \cite{Stefano} is defined by 
\begin{align} \label{eq: pseudo true type 1}
\bvarphi_0^{(m)} \triangleq \arg \max_{\bvarphi^{'}} {\rm E } _g \left[ \log \left (f^{(m)}(\xvec; \bvarphi^{'},\thmtr)\right ) \right]
\end{align}
for type 1, and 
\begin{align} \label{eq: pseudo true type 2}
\begin{split}
\bvarphi_0^{(m)}&\triangleq \bvarphi^{(m)}(\thmz),\\ \thmz &\triangleq\arg \max_{\thm} {\rm E } _g \left[ \log \left (f^{(m)}\left(\xvec;\thm\right)\right ) \right]
\end{split}
\end{align}
for type 2,
for every model index $m \in \mathcal{M}$. 
As described earlier, for type 2 we intend to estimate $\bvarphi^{(m)}(\thm)$. Therefore, the concept of misspecified (MS) unbiasedness of a function of $\thm$ (with pseudo-true parameter $\thmz$) from Definition 3.1 in \cite{Vuong86} and the MCRB for estimation of continuously differentiable function of $\thm$ from Theorem 4.1 in \cite{Vuong86}, are used here.
The MSE from (\ref{MSE phi tau}) can be rewritten as:
\begin{align}\label{eq:9}
    \textbf{MSE}(m) 
    &= {\rm E } _g \Bigg [ \left (\textbf{e}(m)-\bvarphi^{(m)}_0+\bvarphi^{(m)}_0 \right ) \nonumber\\
    &\cdot \left (\textbf{e}(m)-\bvarphi^{(m)}_0+\bvarphi^{(m)}_0 \right )^T \Bigg ] \nonumber \\
&= \left (\bvarphi - \bvarphi^{(m)}_0 \right ) \left (\bvarphi - \bvarphi^{(m)}_0 \right )^T \nonumber \\
&+ \left(\bvarphi - \bvarphi^{(m)}_0 \right) {\rm E } _g \left [ \left(\bvarphi^{(m)}_0 -\hat{\bvarphi}^{(m)} \right)^T \right ] \nonumber \\
 &+ {\rm E } _g \left [ \left(\bvarphi^{(m)}_0 -\hat{\bvarphi}^{(m)} \right )  \right] \left (\bvarphi - \bvarphi^{(m)}_0 \right)^T \nonumber \\
 &+ {\rm E } _g \left [ \left (\hat{\bvarphi}^{(m)} - \bvarphi^{(m)}_0 \right) \left (\hat{\bvarphi}^{(m)} - \bvarphi^{(m)}_0 \right )^T \right ]
\end{align}
in which the first equality is obtained by adding and subtracting the deterministic pseudo-true parameter $\bvarphi_0^{(m)}$. Assuming that $\hat{\bvarphi}^{(m)}$ is a MS-unbiased estimator of $\bvarphi^{(m)}_0$ \cite{Vuong86}, i.e.
\begin{align}
    {\rm E } _g \left[\hat{\bvarphi}^{(m)}-\bvarphi^{(m)}_0\right] =\textbf{0}_s,
\end{align}
the mixed terms in (\ref{eq:9}) vanish. Thus, we obtain:
\begin{equation}\begin{split} \label{eq:10} 
   \textbf{MSE}(m) &=  \left(\bvarphi - \bvarphi^{(m)}_0\right)\left(\bvarphi - \bvarphi^{(m)}_0\right)^T \\
   &+ {\rm E } _g  \left[\left(\hat{\bvarphi}^{(m)} - \bvarphi^{(m)}_0\right) \left(\hat{\bvarphi}^{(m)} - \bvarphi^{(m)}_0\right)^T \right].
\end{split}
\end{equation}
For minimization of the MSE bound w.r.t. $m \in \mathcal{M}$, we define the following scalar function as a measure of performance:
\begin{equation} \label{trace MSE}
\begin{split}
& tr\left(\mathbf{W}\cdot \textbf{MSE}(m)\right) =\left(\bvarphi - \bvarphi^{(m)}_0\right)^T \mathbf{W} \left(\bvarphi - \bvarphi^{(m)}_0\right)\\
&+{\rm E } _g  \left[\left(\hat{\bvarphi}^{(m)} - \bvarphi^{(m)}_0\right)^T \mathbf{W} \left(\hat{\bvarphi}^{(m)} - \bvarphi^{(m)}_0\right)  \right],\\
\end{split}
\end{equation}
where $\mathbf{W} \in \mathbb{R}^{s\times s}$ is a positive-definite weight matrix. For simplicity, we assume $\mathbf{W}=\Imat_s$ and the scalar function in (\ref{trace MSE}) resides to
\begin{align}
tr\left(\textbf{MSE}(m)\right) = \left \| \bvarphi - \bvarphi^{(m)}_0 \right \|^2 + {\rm E } _g \left[\left \| \hat{\bvarphi}^{(m)} - \bvarphi^{(m)}_0 \right \|^2 \right ].
\end{align}
By substituting the MCRB from Theorem 4.1 in \cite{Vuong86} for the second term, the trace of the MSE with order $m$ is bounded by:
\begin{equation}\label{eq:11}
        tr(\textbf{MSE}(m)) \geq B^{(m)},
\end{equation}
where
\begin{equation} \label{eq: Bm}
    \begin{split}
\text{Type 1:}\\
& B^{(m)} \triangleq 
        \left \| \bvarphi - \bvarphi^{(m)}_0 \right \|^2 + tr \left( \textbf{MCRB}\left(\bvarphi^{(m)}_0\right)\right),
\end{split}
\end{equation}
\begin{equation} \label{eq: Bm 2}
    \begin{split}
\text{Type 2:}\\ B^{(m)} &\triangleq 
        \left \| \bvarphi - \bvarphi^{(m)}_0 \right \|^2 \\
&+ tr \left( \frac{\diff \bvarphi_0^{(m)}}{\diff \thmz}\textbf{MCRB}\left(\thmz\right) \left (\frac{\diff \bvarphi_0^{(m)}}{\diff \thmz}\right )^T \right),
    \end{split}
\end{equation}
and the MCRB matrices are defined by (see \cite[Eq. (5)]{Stefano})
\begin{align} \label{eq: MCRB type 1}
    \begin{split}
\textbf{MCRB}(\bvarphi_0^{(m)}) & \triangleq \mathbf{A}^{-1}_{\bvarphi^{(m)}_0} \mathbf{B}_{\bvarphi^{(m)}_0} \mathbf{A}^{-1}_{\bvarphi^{(m)}_0},\\
l^{(1)}(\bvarphi^{(m)}_0) &\triangleq \log \left (f^{(m)}(\xvec; \bvarphi^{(m)}_0,\thmtr)\right ),\\
\mathbf{A}_{\bvarphi^{(m)}_0}  &\triangleq {\rm E } _g \left[  \frac{\diff^2 l^{(1)}(\bvarphi^{(m)}_0)}{\diff\bvarphi^{(m)}_0\diff^T\bvarphi^{(m)}_0}\right], \\
\mathbf{B}_{\bvarphi^{(m)}_0} &\triangleq {\rm E } _g \left[ \frac{\diff l^{(1)}(\bvarphi^{(m)}_0)}{\diff \bvarphi^{(m)}_0}
\frac{\diff l^{(1)}(\bvarphi^{(m)}_0)}{\diff^T \bvarphi^{(m)}_0}\right], 
    \end{split}
\end{align}
and 
\begin{align} \label{eq: MCRB type 2}
    \begin{split}
\textbf{MCRB}(\thmz) &\triangleq \mathbf{A}^{-1}_{\thmz} \mathbf{B}_{\thmz} \mathbf{A}^{-1}_{\thmz}, \\
l^{(2)}(\thmz) &\triangleq \log \left (f^{(m)}\left(\xvec;\thmz\right)\right ),\\
\mathbf{A}_{\thmz}  &\triangleq {\rm E } _g \left[  \frac{\diff^2 l^{(2)}(\thmz)}{\diff\thmz\diff^T\thmz}\right], \\
\mathbf{B}_{\thmz} &\triangleq {\rm E } _g \left[ \frac{\diff l^{(2)}(\thmz)}{\diff \thmz}
\frac{\diff l^{(2)}(\thmz)}{\diff^T \thmz}\right].
\end{split}
\end{align}

The bounds in (\ref{eq: Bm}) and (\ref{eq: Bm 2}) are composed of the norm squared of the bias and the trace of the lower bound on the covariance of $\hat{\bvarphi}^{(m)}$. 

It is proposed to use the MCRB as a criterion for model selection, i.e.
\begin{align} \label{optimal model order}
\begin{split}
    m^* &= \arg \minm_{m \in \mathcal{M}}  B^{(m)}  \; .
\end{split}
\end{align}
Since $g(\xvec;\bvarphi)$ and $\bvarphi$ are not fully known, the MCRB used in (\ref{optimal model order}) cannot be directly computed, and an approximation of the bound for every $m$ and the unknown parameters is required.
In the following Subsections, we present the estimation procedure of (\ref{optimal model order}) for both types.
\subsection{Type 1 Procedure}
As described in Subsection \ref{ssec:NBE:pre}, the model parameters $\thm$ are pre-determined using a training set $\{\xvec_t,\bvarphi_t\}_{t=1}^T$ of size $T$:
\begin{align}
    \thmtr \triangleq \frac{1}{T} \sum_{t=1}^T \arg \max_{\thm} \Big \{ f^{(m)}(\xvec_t; \bvarphi_t,\thm) \Big \}.
\end{align}
Regarding the pseudo-true parameters 
vector $\bvarphi^{(m)}_0$, it can be approximated by the ML estimator 
\begin{align} \label{eq: pseudo-true approx}
\hat{\bvarphi}^{(m)}_{ML} = \arg \max_{\bvarphi} 
    \Big \{ f^{(m)}\left(\xvec; \bvarphi,\thmtr \right) \Big \},
\end{align} by its convergence in distribution (see \cite[Eq. (11)]{Stefano}). If the true PDF is known is some sense, up to unknown nuisance parameters, than the MCRB can be approximated by $\textbf{MCRB}(\hat{\bvarphi}^{(m)}_{ML})$. Otherwise, the sample MCRB 
\begin{align}
    \begin{split}
\Cmat(\bvarphi^{(m)})&\triangleq \hat{\Amat}^{-1} (\bvarphi^{(m)}) \hat{\Bmat}(\bvarphi^{(m)}) \hat{\Amat}^{-1} (\bvarphi^{(m)}),\\
\hat{\Amat}(\bvarphi^{(m)})&\triangleq\left[  \frac{\diff^2 l^{(1)}(\bvarphi^{(m)})}{\diff\bvarphi^{(m)}\diff^T\bvarphi^{(m)}}\right],\\
\hat{\Bmat}(\bvarphi^{(m)}) &\triangleq \left[ \frac{\diff l^{(1)}(\bvarphi^{(m)})}{\diff \bvarphi^{(m)}}
\frac{\diff l^{(1)}(\bvarphi^{(m)})}{\diff^T \bvarphi^{(m)}}\right]
\end{split}
\end{align} at the ML estimator introduces a consistent estimate of the MCRB (see \cite[Eq. (16)]{Stefano3})
\begin{align}
\begin{split}
\Cmat\left(\hat{\bvarphi}^{(m)}_{ML} \right) \xrightarrow[]{a.s.} \textbf{MCRB}(\bvarphi_0^{(m)}).
\end{split}
\end{align}
For approximation of the true parameter vector $\bvarphi$, estimation of the highest model order pseudo-true parameter vector $\bvarphi^{(M)}_0$ by $\hat{\bvarphi}^{(M)}_{ML}$ will be used. For cases that the models are nested and the true model is inside the class of models, $\bvarphi=\bvarphi^{(M)}_0$ is obtained.
The resulted MCRB-based model selection criterion estimation is given by 
\begin{align} \label{eq: Bm est 1}
    \hat{B}^{(m)} = \left \| \hat{\bvarphi}^{(M)}_{ML} - \hat{\bvarphi}^{(m)}_{ML}\right \|^2 + tr \left( \textbf{MCRB}\left(\hat{\bvarphi}^{(m)}_{ML}\right)\right)
\end{align}
\subsection{Type 2 Procedure}
Regarding the pseudo-true parameters vector $\thmz$, it can be approximated by the ML estimator 
\begin{align} \label{eq: ML type 2}
    \thML = \arg \max_{\thm} \Big \{ f^{(m)}\left(\xvec;\thm\right) \Big \}.
\end{align}
If the true PDF is known is some sense, up to unknown nuisance parameters, than the MCRB can be approximated by $\textbf{MCRB}(\thML)$. Otherwise, the sample MCRB 
\begin{align}
    \begin{split}
\Cmat(\thm)&\triangleq \hat{\Amat}^{-1} (\thm) \hat{\Bmat}(\thm) \hat{\Amat}^{-1} (\thm),\\
\hat{\Amat}(\thm)&\triangleq\left[  \frac{\diff^2 l^{(2)}(\thm)}{\diff\thm \diff^T\thm}\right],\\
\hat{\Bmat}(\thm) &\triangleq \left[ \frac{\diff l^{(2)}(\thm)}{\diff \thm}
\frac{\diff l^{(2)}(\thm)}{\diff^T \thm}\right]
\end{split}
\end{align} at the ML estimator introduces a consistent estimate of the MCRB (see \cite[Eq. (16)]{Stefano3})
\begin{align}
\begin{split}
\Cmat\left(\thML \right) \xrightarrow[]{a.s.} \textbf{MCRB}(\thmz).
\end{split}
\end{align}
For approximation of the true parameter vector $\bvarphi$, estimation of the highest model order pseudo-true parameter vector $\bvarphi^{(M)}_0$ by $\bvarphi^{(M)}\left(\hat{\btheta}^{(M)}_{ML}\right)$ will be used. For cases that the models are nested and the true model is inside the class of models, $\bvarphi=\bvarphi^{(M)}_0$ is obtained.
The resulted MCRB-based model selection criterion estimation is given by 
\begin{align} \label{eq: B 2 est}
\begin{split}
     \hat{B}^{(m)} &= \left \| \bvarphi^{(M)}\left(\hat{\btheta}^{(M)}_{ML}\right) - \bvarphi^{(m)}\left(\hat{\btheta}^{(m)}_{ML}\right) \right \|^2 \\
&+ tr \left( \frac{\diff \bvarphi_0^{(m)}}{\diff \thML}\textbf{MCRB}\left(\thML\right) \left (\frac{\diff \bvarphi_0^{(m)}}{\diff \thML}\right )^T \right)
\end{split}
\end{align}

After determining the model which minimizes the MCRB, the ML estimator for the parameter-of-interest can be obtained as follows:
{\small
\begin{align} \label{eq: ML types}
\begin{split}
    &\text{Type 1:}\\
     &\hat{ \bvarphi}^{(m*)}_{ML} = \arg \max_{\bvarphi} 
    \Big \{ f^{(m*)}\left(\xvec; \bvarphi,\boldsymbol{\theta}^{(m*)}_{tr} \right) \Big \} ,\\
    &\text{Type 2:}\\
   &\hat{ \bvarphi}^{(m*)}_{ML} = \bvarphi^{(m*)} \left (\arg \max_{\boldsymbol{\theta}^{(m*)}} 
    \Big \{ f^{(m*)}\left(\xvec; \boldsymbol{\theta}^{(m*)} \right) \Big \} \right) .
\end{split}
\end{align}
}

\section{Applications}
\label{Applications}
\indent In this section, we provide two signal processing problems which involve parameter estimation based on the two types of problems presented in Section \ref{sec:MCRB for model order selection}. In Subsection \ref{sec: DOA_Estimation}, we address the problem of DOA estimation by a radar system, in the presence of unknown clutter. This problem can be stated as the type 1. In Subsection \ref{sec: spectrum estimation}, we address the problem of parametric spectrum estimation, which can be stated as the type 2.
\subsection{DOA Estimation Under Unknown Clutter/Interference} \label{sec: DOA_Estimation}
In DOA estimation problems in radar systems, the spatial covariance matrix of the sensor measurements is frequently unknown, while it can be estimated from other target-free measurements, such as adjacent range-Doppler cells. Kelly's detection algorithm \cite{Kelly} used a set of signal-free data vectors (referred secondary) to estimate the covariance matrix, and to derive a likelihood ratio test (LRT) of signal presence in another data vector (primary). Another generalized likelihood ratio test (GLRT) detection algorithms that use similar data sets are presented in \cite{Maio}. Consider the following basic single-input multiple-output (SIMO) model:
\begin{equation}
    \xvec = \textbf{a}(\psi) s + \textbf{w},
\end{equation}
where $s\in \mathbb{C}$ is an unknown signal and $\psi \in \mathbb{R}$ is the unknown DOA. The function $\textbf{a}:\mathbb{R} \rightarrow \mathbb{C}^N$ is a known steering vector function, and $\textbf{w} \sim N^C(\textbf{0},\textbf{R})$ denotes the additive colored Gaussian noise. The ML estimator of the DOA is given by \cite{van2004optimum}: 
\begin{equation}
    \hat{\psi}_{ML} = \arg \max_{\psi} \Bigg \{ \frac{\left | \xvec^H \textbf{R}^{-1} \textbf{a}(\psi) \right |^2}{\textbf{a}^H(\psi) \textbf{R}^{-1}  \textbf{a}(\psi)} \Bigg \},
\end{equation}
which depends on the covariance matrix $\textbf{R}$. The covariance matrix, $\textbf{R}$, can be estimated using target-free samples before detection or DOA estimation. The ML estimator of the covariance matrix of the vector $\xvec \in \mathbb{C}^N$ over $T$ i.i.d. samples $\xvec_t, t=1,\dots,T$ is given by:
\begin{equation}\label{Rx est}
    \hat{\textbf{R}}_\xvec = \frac{1}{T}\sum_{t=1}^T \xvec_t\xvec_t^H.
\end{equation}

This application is focused on analyzing the target DOA estimation problem, using training samples to estimate the covariance matrix $\textbf{R}_\xvec$ under a misspecified model regarding the second order statistics. An approximation of the MCRB is used here to obtain a model selection criterion (\ref{optimal model order}), as previously described.
As defined in \cite{Kelly}, data samples are separated into two data sets.
\subsubsection{Adjacent cells training set}
   The first training set is composed of noise and/or interference and/or clutter, without target. It consists of $T$ samples, dedicated to estimate the covariance matrix of $\xvec$, which are sampled under the statistical model:
\begin{equation} \label{eq: secondary set 1}
    \xvec_t = \textbf{c}_t + \vvec_t, \indent t=1,\dots,T, 
\end{equation}
where $\left \{\vvec_t\right\} \in \mathbb{C}^N$ is an i.i.d. sequence of complex Gaussian noise vector with covariance matrix $\sigma^2\textbf{I}_N$, and $\left\{\textbf{c}_t\right\} \in \mathbb{C}^N$ is an i.i.d. sequence of complex random interference/clutter vector.
Note that $\{\textbf{c}_t\}$ and $\{\vvec_t\}$ are assumed statistically independent. The covariance matrix of $\xvec_t$ is given by
\begin{equation}
    \textbf{R}_{\xvec} = \textbf{R}_{\textbf{c}} + \sigma^2 \textbf{I}_N,
\end{equation}
where $\textbf{R}_{\textbf{c}}\in\mathbb{C}^{N\times N}$ is the clutter sequence covariance matrix.
\subsubsection{Target data set}
The second data set is of target and noise and/or interference and/or clutter. It is composed of $D$ samples defined by:
\begin{equation}\begin{split}
     \xvec_{d,j} &= \textbf{a}(\psi_d) s_{d,j} + \textbf{c}_{d,j} + \vvec_{d,j}, \indent j=1,\dots,D\\
     \mathbf{X}_d &= \left[\xvec_{d,1},\dots,\xvec_{d,D}\right],\quad \svec_d = \left[s_{d,1},\dots,s_{d,D}\right]^T,
\end{split}
\end{equation}
where $\psi_d, \big \{ s_{d,j} \big \}_{j=1}^{D}$ denote the true signal DOA, and the complex signal intensity, respectively, and $\textbf{c}_{d,j}, \vvec_{d,j}$ share the same statistics as $\textbf{c}_t, \vvec_t$, that is the same mean and covariance. Note that $\psi_d, \big \{ s_{d,j} \big \}_{j=1}^{D},$ and $\textbf{R}_\xvec$ are deterministic unknown and are to be estimated. The parameter vector is defined by
\begin{equation}
    \bvarphi = \left[\psi, \Re \left [s_j\right], \Im \left [s_j\right] \right]^T,
\end{equation}
where the parameter of interest is the DOA $\psi$. The estimated model parameters $\thm$ are the assumed covariance matrix $\textbf{R}_{\xvec,m}$ of index $m$, meaning $\thm\triangleq\textbf{R}_{\xvec,m}$
The true data PDF is given by
\begin{equation} \label{eq: true data PDF}
    g(\xvec_{d,j};\bvarphi_d) : \xvec_{d,j} \sim  \mathcal{N}^\mathcal{C}(\textbf{a}(\psi_d) s_{d,j}, \textbf{R}_\xvec),
\end{equation}
and the assumed PDF is
\begin{equation} \label{eq: assumed data PDF 1}
    f^{(m)}(\xvec_{d,j}; \bvarphi,\thm) : \xvec_{d,j} \sim \mathcal{N}^\mathcal{C}(\textbf{a}(\psi) s_{j},\textbf{R}_{\xvec,m}).
\end{equation}

A common decomposition of $\textbf{R}_\xvec$ includes separating it to noise subspace and interference and noise subspace using singular value decomposition (SVD) \cite{van2004optimum}:
\begin{equation}
    \textbf{R}_\xvec = \textbf{U} \Lambda \textbf{U}^H,
\end{equation}
where $\textbf{U}$ is a unitary matrix of eigenvectors, and diagonal eigenvalues matrix $\Lambda$, with the eigenvalues in $\Lambda$ arranged in descending order \cite{van2004optimum}. 
The assumed model of index $m=0,\dots,N-1$ is defined by decomposition of $\textbf{R}_\xvec$ into a interference and noise subspace of $m$ degree  and a noise subspace of $N-m$ degree, to form the assumed covariance matrix 
$\textbf{R}_{\xvec,m}$.
 It is pre-determined using the first training set. The martix $\thmtr \triangleq \Rxm$ estimate would be to compute the SVD of the r.h.s of (\ref{Rx est}) to obtain $\hat{\textbf{U}}$ and $ \hat{\Lambda}=\Diag(\hat{\boldsymbol{\lambda}}),  \hat{\boldsymbol{\lambda}} = [\hat{\lambda}_1,\dots,\hat{\lambda}_N]^T$, and then estimate the noise variance by:
\begin{equation}
\sm = \frac{1}{N-m} \sum_{i=m+1}^N \hat{\lambda}_i.    
\end{equation}
Using the common seperation into subspaces of $\textbf{R}_{\xvec,m}$, its estimate is defined by:
\begin{align}\label{Rxm}
\begin{split}
   & \Rxm = \hat{\textbf{U}} \hat{\Lambda}_m \hat{\textbf{U}}^H, \indent  \hat{\Lambda}_m = \Diag(\hat{\boldsymbol{\lambda}}_m),\\
   & \hat{\boldsymbol{\lambda}}_m = [\hat{\lambda}_1,\dots,\hat{\lambda}_m,\sm,\dots,\sm]^T,
   \end{split}
\end{align}
which enforces noise variance of $\sm$ and $m$ larger eigenvalues $\hat{\lambda}_i, i=0,\dots,m$, corresponding to interference and noise subspace.
Note that the model is misspecified due to error in covariance matrix estimate, $\Rxm$, which is computed using the first training set, and assumed to be deterministic.
The corresponding assumed PDF is given by 
\begin{equation} \label{eq: assumed data PDF}
    f^{(m)}(\xvec_{d,j}; \bvarphi,\thmtr) = \mathcal{N}^\mathcal{C}(\textbf{a}(\psi) s_{j},\Rxm).
\end{equation}

\subsubsection{MCRB based criterion}
To estimate the DOA, $\psi_d$, we exploit the structure of the covariance matrix $\Rxm$. The candidate models are defined such that the covariance matrix is composed of $m = 0,\dots,N-1$ interferences and noise. We intend to minimize the MSE, but not necessarily select the true noise model.
As described in Section \ref{sec:MCRB for model order selection}, MCRB-based model selection criterion is used (\ref{optimal model order}). Thus, we derive the MCRB in the following.
The pseudo-true parameter vector $\bvarphi_0^{(m)}$ is given by (\ref{eq: pseudo true type 1}):
\begin{align} \label{eq: pseudo true parameter vecotr theta}
\begin{split}
    \bvarphi_0^{(m)} &= \arg \max_{\bvarphi} {\rm E } _g \left[ \log \left (f^{(m)}(\mathbf{X}_{d}; \bvarphi,\Rxm)\right ) \right].
\end{split}
\end{align}
Looking at the log-likelihood function of the assumed model:
\begin{align}
\begin{split}    
    &\log \left (f^{(m)}(\mathbf{X}_{d}; \bvarphi,\Rxm)\right)  = -D\cdot \log det\left(\pi \Rxm \right) \\
    &-\sum_{j=1}^{D} \left(\xvec_{d,j} - \textbf{a}(\psi) s_{j}\right)^H \Rxminv \left(\xvec_{d,j} - \textbf{a}(\psi) s_{j} \right),
    \end{split}
\end{align}
and obtaining the expectation w.r.t. the true model results in:
\begin{align} \label{eq: log likelihood expectation}
\begin{split}
&{\rm E } _g \left[ \log \left (f^{(m)}(\mathbf{X}_{d}; \bvarphi,\Rxm)\right) \right] =  -D\cdot \log det\left(\pi \Rxm \right) \\ 
& -\sum_{j=1}^{D}\left [ {\rm E } _g \left[  \xvec^H_{d,j} \Rxminv \xvec_{d,j}\right] -2  \Re \left [s^*_{d,j}\textbf{a}^H(\psi_d) \Rxminv \textbf{a}(\psi) s_{j}\right]\right] \\
& -\sum_{j=1}^{D}\left [ \textbf{a}^H(\psi) \Rxminv \textbf{a}(\psi) |s_{j}|^2 \right].
\end{split}
\end{align}
Substituting (\ref{eq: log likelihood expectation}) in (\ref{eq: pseudo true parameter vecotr theta}), and using the fact that the first two terms in the r.h.s. of (\ref{eq: log likelihood expectation}) are independent on $\bvarphi$, one obtains:
\begin{align} \label{eq: 49}
\begin{split}
    \bvarphi_0^{(m)} &= \arg \max_{\bvarphi}  \sum_{j=1}^{D} 2\Re \left [s^*_{d,j}\textbf{a}^H(\psi_d) \Rxminv \textbf{a}(\psi) s_{j}\right]\\
    & -\sum_{j=1}^{D}\textbf{a}^H(\psi) \Rxminv \textbf{a}(\psi) |s_{j}|^2. 
\end{split}
\end{align}
First derivative of the r.h.s. of (\ref{eq: log likelihood expectation}) w.r.t. $s_j$ results in:
\begin{align}
\begin{split}
    &\frac{\partial {\rm E } _g \left[ \log \left (f^{(m)}(\mathbf{X}_{d}; \bvarphi,\Rxm)\right) \right]}{\partial s_j}  =\\
    & s^*_{d,j}\textbf{a}^H(\psi_d) \Rxminv \textbf{a}(\psi)
    - s^*_j \textbf{a}^H(\psi) \Rxminv\textbf{a}(\psi),
\end{split}
\end{align}
and imposing equality to zero result in the pseudo-true signal 
\begin{equation} \label{eq: pseudo true signal}
    s_{j,0}^{(m)} = \frac{ \textbf{a}^H(\psi) \Rxminv \textbf{a}(\psi_d)}{\textbf{a}^H(\psi) \Rxminv\textbf{a}(\psi)} s_{d,j}.
\end{equation}
Substituting (\ref{eq: pseudo true signal}) in (\ref{eq: 49}) and considering the first element of $\bvarphi_0^{(m)}$ results in:
\begin{align} \label{eq: 52}
    \begin{split}   \left[\bvarphi_0^{(m)}\right]_1 &=  \arg \max_{\psi} \Biggl\{ \|\svec_d\|^2  \cdot \frac{\left |\textbf{a}^H(\psi_d) \Rxminv \textbf{a}(\psi) \right |^2}{\textbf{a}^H(\psi) \Rxminv \textbf{a}(\psi)}  \Biggl\} \\
    &=\arg \max_{\psi} \Biggl\{  \frac{\left |\textbf{a}^H(\psi_d) \Rxminv \textbf{a}(\psi) \right |^2}{\textbf{a}^H(\psi) \Rxminv \textbf{a}(\psi)}  \Biggl\} = \psi_d,
    \end{split}
\end{align}
where the second equality is obtained because the signal intensity $\|\svec_d\|^2 = \sum_{j=1}^{D}\left|s_{d,j}\right|^2$ is independent of $\psi$, and the third by the Cauchy-Schwarz equality condition. Substituting (\ref{eq: 52}) in (\ref{eq: pseudo true signal}) verifies that the pseudo-true parameter vector $\bvarphi_0^{(m)}$ coincides with the true parameter vector:
\begin{equation} \label{eq: 43}
    \bvarphi_0^{(m)} = \bvarphi_d, \quad m= 0,\dots,N-1.
\end{equation}
Thus, the norm squared of the bias in the r.h.s. of (\ref{eq: Bm}) equals zero, and the MSE bound is dependent only on the covariance term. 
Note that the signal is considered a nuisance parameter.
For the problem presented by the true and assumed PDFs in (\ref{eq: true data PDF}),(\ref{eq: assumed data PDF}), the MCRB matrices (\ref{eq: MCRB type 1}) result in:
\begin{align}\label{eq: B DOA}
\begin{split}
 \mathbf{B}_{\bvarphi^{(m)}_0}
&= 2 D
\begin{bmatrix}
a_j & \Re \left[d_j\right] & -\Im \left[d_j\right] \\
\Re \left[d_j\right] & c_j & 0\\
-\Im \left[d_j\right] & 0 & c_j
\end{bmatrix},
\end{split}
\end{align}
where 
\begin{align}
\begin{split}
    a_j &= \dot{\textbf{a}}^H(\psi_d) \Rxminv \textbf{R}_\xvec \Rxminv \dot{\textbf{a}}(\psi_d) \frac{\|\svec_d\|^2}{D} ,\\
    c_j &= \textbf{a}^H(\psi_d) \Rxminv \textbf{R}_\xvec \Rxminv \textbf{a}(\psi_d),\\
    d_j &= \dot{\textbf{a}}^H(\psi_d)  \Rxminv \textbf{R}_\xvec \Rxminv \textbf{a}(\psi_d) \frac{1}{D} \sum_{j=1}^{D} s^*_{d,j},
\end{split}
\end{align} 
and
\begin{equation}\label{eq: A DOA}
\mathbf{A}_{\bvarphi^{(m)}_0} =-2 D  \begin{bmatrix}  b_j   &
\Re \left [e_j \right]  &   -\Im \left [e_j\right]\\
  \Re \left [e_j\right]& f_j & 0  \\
  -\Im [e_j] & 0 & f_j
\end{bmatrix},
\end{equation}
where
\begin{align} \begin{split}
b_j &=\dot{\textbf{a}}^H(\psi_d) \Rxminv \dot{\textbf{a}}(\psi_d) \frac{\|\svec_d\|^2}{D} ,\\
e_j &=\dot{\textbf{a}}^H(\psi_d)  \Rxminv \textbf{a}(\psi_d) \frac{1}{D} \sum_{j=1}^{D} s^*_{d,j},\\
f_j &= \textbf{a}^H(\psi_d) \Rxminv \textbf{a}(\psi_d).
\end{split}
\end{align}

For presentation simplicity, we suggest to derive the MCRB s.t. $\sum_{j=1}^{D} s_{d,j}=0$ is obtained. It can be shown that $d_j=e_j=0$, and $\mathbf{A}_{\bvarphi^{(m)}_0} ,\mathbf{B}_{\bvarphi^{(m)}_0}$ are diagonal.

Substituting (\ref{eq: 43}), (\ref{eq: B DOA}), and (\ref{eq: A DOA}) in (\ref{eq: Bm}) results in the bound for the parameter of interest $\psi$

\begin{align} \label{eq: B m DOA under clutter}
\begin{split}
B^{(m)} &=\left [\textbf{ MCRB}(\bvarphi_0^{(m)})\right]_{1,1}\\
&= \frac{D}{ \|\svec_d\|^2}\cdot \frac{\dot{\textbf{a}}^H(\psi_d) \Rxminv \textbf{R}_\xvec \Rxminv \dot{\textbf{a}}(\psi_d)}{\left(\dot{\textbf{a}}^H(\psi_d) \Rxminv \dot{\textbf{a}}(\psi_d)\right)^2 }.
\end{split}
\end{align}

Looking at the bound, we see that it is dependent of $\psi_d,\|\svec_d\|^2, \textbf{R}_\xvec$, which are unknown. To face this issue, as suggested in (\ref{eq: pseudo-true approx}), we substitute the first by the maximum likelihood estimator
\begin{equation} \label{eq: DOA under clutter ML estimations}
    \hat{\psi}^{(m)}_{ML} = \arg \max_{\psi} \frac{\left \| \mathbf{X}^H_{d}\Rxminv\textbf{a}(\psi) \right \|^2}{\textbf{a}^H(\psi) \Rxminv \textbf{a}(\psi)} .
\end{equation}
The true signal intensity $\|\svec_d\|^2$ in the denominator of (\ref{eq: B m DOA under clutter}) is independent on the model $m$, thus it is ignored in the selection process.
The matrix $\textbf{R}_\xvec$ is estimated by the highest order $M=N-1$ assumed covariance matrix $\RxM$, which coincides with $\hat{\textbf{R}}_{\xvec}$ in (\ref{Rx est}), an unbiased estimator.

The MCRB-based criterion is given by substituting (\ref{eq: DOA under clutter ML estimations}) and $\RxM$ in (\ref{eq: B m DOA under clutter}):
\begin{align}
 \hat{B}^{(m)} = \frac{D}{ \|\svec_d\|^2}\cdot \frac{\dot{\textbf{a}}^H(\psimML) \Rxminv \RxM \Rxminv \dot{\textbf{a}}(\psimML)}{\left(\dot{\textbf{a}}^H(\psimML) \Rxminv \dot{\textbf{a}}(\psimML)\right)^2 }. 
\end{align}
and the estimator of $\varphi$ based on the selected model order $m^*$ is
\begin{equation}
    \hat{\psi}^{(m^*)} =\hat{\psi}^{(m^*)}_{ML}.
\end{equation}

\subsection{Spectrum Estimation - Auto-Regressive Model}\label{sec: spectrum estimation}

Spectrum estimation of a random stationary process is a classical signal processing problem. The two main approaches are parameteric and non-parametric estimation. A simple model which is generated with an innovation process and a digital filter is the auto-regressive moving average (ARMA) model. The estimation of its spectrum and Cram\'{e}r-Rao type bounds have been investigated for Gaussian models \cite{porat1994digital}. In this Subsection, we analyze ARMA$(p,q)$ models under unknown number of parameters $p,q$, and a class of AR$(m)$ models which are used to approximate them. The candidate models for estimation of the spectral density will be a collection of AR$(m)$ models, defined with varying number of AR parameters, $m \in \mathcal{M}$. Specifically, $\mathcal{M}=\{ 1,\dots,M \}$, with $M > p+q$. 

The first part is dedicated to deriving bounds for the MSE, defined in (\ref{eq: Bm 2}), for spectral estimation with this class of models. The bounds derived using the MCRB demonstrate the large estimation errors due to misspecification, when the true model includes a moving average (MA) part.

The second part is aimed at the task of model selection and spectral density estimation, which is considered using the MCRB-based criterion in (\ref{optimal model order}).
\subsubsection{ARMA model}
The true ARMA$(p,q)$ model is defined by:
\begin{equation}\label{eq: Arma}
    x_t = -\sum_{k=1}^p a_k x_{t-k} + \sum_{k=0}^q b_k u_{t-k},
\end{equation}
with $\textbf{a}_t = [a_p,\dots, a_1]^T,\textbf{b}_t = [b_q,\dots,b_0]^T$ being the true coefficients, $\{u_t\}$ with zero mean and $\sigma_u^2$ variance is the innovation process, and the parameter vector of estimation is
\begin{equation}
\boldsymbol{\theta}_{t} = [\sigma_u^2, \textbf{a}_t^T,\textbf{b}_t^T]^T.
\end{equation}
The true spectral density function is given by \cite{porat1994digital}:
\begin{equation}
    S(e^{j\omega}) = \frac{\sigma_u^2 |b(e^{j\omega})|^2}{|a(e^{j\omega})|^2} = \frac{\sigma_u^2 |\sum_{k=0}^q b_k e^{-j\omega k}|^2}{|\sum_{k=0}^p a_k e^{-j\omega k}|^2},\quad a_0 = 1,
\end{equation}
and the parameter of interest is given as a function of frequency $\omega \in [0,2\pi)$:
\begin{equation}
\varphi(\omega) = \log S(e^{j\omega}).    
\end{equation}
\subsubsection{AR model}
The assumed AR($m$) Gaussian model can be defined similarly to (\ref{eq: Arma}) with $m$ coefficients. In this case, the parameter vector of estimation is:
\begin{equation}
\boldsymbol{\theta}^{(m)} = [\sigma_u^2, \textbf{a}_m^T]^T, \indent \textbf{a}_m=[a_m,\dots,a_1]^T,
\end{equation}
with the corresponding:
\begin{align} \begin{split}
     S^{(m)}(e^{j\omega}) &= \frac{\sigma_u^2}{|a^{(m)}(e^{j\omega})|^2} = \frac{\sigma_u^2}{|\sum_{k=0}^m a_k e^{-j\omega k} |^2}, \\ 
     \varphi ^{(m)}(\omega) &= \log S^{(m)}(e^{j\omega})
\end{split}
\end{align}
\subsubsection{Log-likelihood function}
To derive the likelihood function under the assumed model, we first consider the likelihood function as presented in \cite{porat1994digital}. The samples vector of size $T$ is: $\xvec_T = [x_0,\dots,x_{T-1}]^T$, with $T \geq 3m+1$. According to \cite[Theorem 6.4.]{porat1994digital}, the assumed log likelihood function can be described using a sufficient statistic:
\begin{equation}
    T(\xvec_T) = [x_0,\dots,x_{m-1},x_{T-m},\dots,x_{T-1},\hat{r}_0,\dots,\hat{r}_m],
\end{equation}
with $\hat{r}_i, i=0,\dots,m$ being the biased sample covariances
\begin{align}
    \hat{r}_m \triangleq \frac{1}{T} \sum_{t=0}^{T-m-1} x_{m+t}x_t .   
\end{align}
Using this statistic, the assumed log likelihood function is derived:
{\small
\begin{equation} \begin{split} \label{log likelihood m}
    &\log \left( f^{(m)}(\xvec_T; \boldsymbol{\theta}^{(m)})\right) = - \frac{T}{2} \log(2\pi \sigma_u^2) +\frac{1}{2} \sum_{n=1}^m n \log(1-K_n^2) \\
    &-\frac{1}{2\sigma_u^2}[\textbf{a}_m^T, 1] \Big [T\hat{\textbf{R}}_{m+1} - \Xmat_1^T \Xmat_1 - \Xmat_2^T \Xmat_2\Big] [\textbf{a}_m^T, 1]^T,
    \end{split}
\end{equation}
}
where the partial correlation coefficient is 
\begin{align}
\begin{split}
    K_n &\triangleq \frac{{\rm E } _g \left[u_{t,n}\bar{u}_{t-n-1,n}\right]}{\sqrt{{\rm E } _g \left[u^2_{t,n}\right]} \sqrt{{\rm E } _g \left[\bar{u}^2_{t-n-1,n}\right]}},\\
    u_{t,n} &\triangleq \sum_{k=0}^n\alpha_{n,k}x_{t-k}, \quad \alpha_{n,0}=1,\\
    & \bar{u}_{t,n} \triangleq \sum_{k=0}^n \bar{\alpha}_{n,k} x_{t+k},\quad \bar{\alpha}_{n,0}=1,
\end{split}
\end{align} 
$\{-\alpha_{n,k}, 1\leq k \leq n\}, \{-\bar{\alpha}_{n,k}, 1\leq k \leq n\}$ are the coefficients of the $n$th-order best linear predictors of $x_t$ from $\{x_{t-k},1\leq k \leq n\}, \{x_{t+k},1\leq k \leq n\}$, respectively.
 The sample covariance matrix elements are given by
 \begin{align} \label{eq: R p+1}
\begin{split}
\left [\hat{\textbf{R}}_{m+1}\right]_{i,j} &= \frac{1}{T} \Xmat^T \Xmat =\hat{r}_{|i-j|},\quad i,j=1,\dots,m+1,\\
\end{split}
\end{align}
where the $(T+m)\times (m+1)$ measurements matrix $\Xmat$ is (see \cite[Eq. (6.1.4)]{porat1994digital}) 
 {\normalsize
\begin{align} \label{eq: Xmat}
\begin{split}
\Xmat &= 
\begin{bmatrix}
    x_0 & 0 & \dots & 0\\
    x_1 & x_0 & \dots & \dots\\
    \dots & \dots & \dots & 0\\
    x_{m} & x_{m-1} & \dots & x_0\\
    \dots & \dots & \dots & \dots\\
    x_{T-1} & x_{T-2} & \dots & x_{T-m-1}\\
    0 & x_{T-1} & \dots & x_{T-m}\\
    \dots & 0 & \dots & \dots\\
    \dots & \dots & \dots & \dots\\
    0 & 0 & \dots & x_{T-1}\\
\end{bmatrix},\\
\end{split}
\end{align}
    }
and the following matrices are
\begin{align} \label{Y matrices}
\begin{split}
    \Xmat_1^T &= \begin{bmatrix}
x_0 & x_1 & \dots & x_{m-1}\\
0 & x_0 & \dots & x_{m-2}\\
\dots & \dots & \dots & \dots\\
0 & \dots & 0 & x_0\\
0 & \dots & \dots & 0\\
\end{bmatrix} \in \mathbb{R}^{(m+1) \times m},\\
\Xmat_2^T &= \begin{bmatrix}
x_{T-1} & x_{T-2} & \dots & x_{{T-m}}\\
0 & x_{T-1} & \dots & x_{T-m-1}\\
\dots & \dots & \dots & \dots\\
0 & \dots & 0 & x_{T-1}\\
0 & \dots & \dots & 0\\
\end{bmatrix} \in \mathbb{R}^{(m+1) \times m}.\\
\end{split}
\end{align}
For derivation of the MCRB, we need to differentiate w.r.t. $\boldsymbol{\theta}^{(m)}$. The derivative of the second term in the r.h.s. in (\ref{log likelihood m}) cannot be analytically computed, like ML estimation which involves non linear operations, as described in \cite{porat1994digital}. Nevertheless, this term is negligible as the number of samples $T$ increases. Thus, the log likelihood will be estimated by:
\begin{equation}\label{log likelihood m est}\begin{split}
    &l^{(2)}(\thm)= - \frac{T}{2} \log(2\pi \sigma_u^2)\\
    &-\frac{1}{2\sigma_u^2}[\textbf{a}_m^T, 1] \Big [T\hat{\textbf{R}}_{m+1} - \Xmat_1^T \Xmat_1 - \Xmat_2^T \Xmat_2\Big] [\textbf{a}_m^T, 1]^T.
\end{split}
\end{equation}
First derivative w.r.t. $\sigma_u^2$ yields:
\begin{equation} \label{eq: der 1}
\begin{split}
    &\frac{ \partial l^{(2)}(\thm)}{\partial \sigma_u^2} = -\frac{T}{2 \sigma_u^2}\\
    &+ \frac{1}{2 \sigma_u^4} [\textbf{a}_m^T, 1] \Big [T\hat{\textbf{R}}_{m+1} - \Xmat_1^T \Xmat_1 - \Xmat_2^T \Xmat_2\Big] [\textbf{a}_m^T, 1]^T,
\end{split}    
\end{equation}
and the gradient w.r.t. $\textbf{a}_m$:
{\small
\begin{equation} \begin{split}
    &\frac{d^T l^{(2)}(\thm) }{d \textbf{a}_m} = -\frac{1}{\sigma_u^2} [\textbf{a}_m^T, 1] \Xmat^T \Tilde{\Xmat} + \frac{1}{\sigma_u^2} \textbf{a}_m^T \Big [\Tilde{\Xmat}_1^T \Tilde{\Xmat}_1 + \Tilde{\Xmat}_2^T \Tilde{\Xmat}_2 \Big],
    \end{split}
\end{equation}
}
where the matrices elements are:
\begin{align}
    \begin{split}
\left[\Tilde{\Xmat}\right]_{k,l} &= \left[\Xmat\right]_{k,l},\quad k=1,\dots,T+m, \quad l=1,\dots,m,\\
\left[\Tilde{\Xmat}_1\right]_{k,l} &= \left[\Xmat_1\right]_{k,l},\quad k,l=1,\dots,m,\\
\left[\Tilde{\Xmat}_2\right]_{k,l} &= \left[\Xmat_2\right]_{k,l},\quad k,l=1,\dots,m 
.
    \end{split}
\end{align}
The second derivatives are given by
{\small
\begin{equation} \begin{split}
     &\frac{ \partial^2 l^{(2)}(\thm)}{\partial \sigma_u^2 \partial \sigma_u^2} = \frac{T}{2 \sigma_u^4} \\
     &- \frac{1}{ \sigma_u^6} [\textbf{a}_m^T, 1] \Big [T\hat{\textbf{R}}_{m+1} - \Xmat_1^T \Xmat_1 - \Xmat_2^T \Xmat_2\Big] [\textbf{a}_m^T, 1]^T,
\end{split}
\end{equation}
\begin{equation} \begin{split}
       &\frac{ d^2 l^{(2)}(\thm)}{d \textbf{a}^T_m\partial \sigma_u^2 } = \frac{1}{ \sigma_u^4} [\textbf{a}_m^T, 1] \left [\Xmat^T \Tilde{\Xmat} \right ]
       - \frac{1}{ \sigma_u^4} \textbf{a}_m^T \left [ \Tilde{\Xmat}_1^T \Tilde{\Xmat}_1 + \Tilde{\Xmat}_2^T \Tilde{\Xmat}_2\right],
\end{split}
\end{equation}
\begin{equation} \label{eq: der end}\begin{split}
    &\frac{ d^2 l^{(2)}(\thm)}{d \textbf{a}_m d^T \textbf{a}_m} = 
    - \frac{1}{ \sigma_u^2} \left [\Tilde{\Xmat}^T \Tilde{\Xmat} -\Tilde{\Xmat}_1^T \Tilde{\Xmat}_1 - \Tilde{\Xmat}_2^T \Tilde{\Xmat}_2\right].
    \end{split}
\end{equation}
}
\subsubsection{The MCRB} \label{Sec: MCRB}
To evaluate the MCRB for the assumed AR($m$) model, the pseudo-true parameter vector $\boldsymbol{\theta}^{(m)}_{0}=[\sigma_{u,0}^2,\textbf{a}_{m,0}^T]^T$ is required. As was shown in \cite{Stefano}, the ML estimator converges almost surely (a.s.) to the pseudo-true parameter. For AR models, the ML estimate is asymptotically equivalent to the Yule-Walker estimate. Therefore, Yule-Walker estimation of $\boldsymbol{\theta}^{(m)}$ converges a.s. to the pseudo-true parameter vector, and obtaining the empirical expectation of it with large number of samples $T$ will result in the pseudo-true parameter vector. 
Substituting (\ref{log likelihood m est})-(\ref{eq: der end}) evaluated at $\thmz$ in (\ref{eq: MCRB type 2}) gives the MCRB matrices 
\begin{align} \label{eq: spectrum MCRB matrices}
\begin{split}
    \mathbf{A}_{\thmz}  &= {\rm E } _g   \begin{bmatrix}
\frac{ \partial^2 l^{(2)}(\thmz)}{\partial \sigma_{u,0}^2 \partial \sigma_{u,0}^2} & \frac{ d^2 l^{(2)}(\thmz)}{d \textbf{a}^T_{m,0}\partial \sigma_{u,0}^2 } \\
\frac{ d^2 l^{(2)}(\thmz)}{d \textbf{a}_{m,0}\partial \sigma_{u,0}^2 }  & \frac{ d^2 l^{(2)}(\thmz)}{d \textbf{a}_{m,0} d^T \textbf{a}_{m,0}} \\
\end{bmatrix}, \\
\mathbf{B}_{\thmz} &= {\rm E } _g  
\begin{bmatrix} 
\left(\frac{ \partial l^{(2)}(\thmz)}{\partial \sigma_{u,0}^2}\right)^2 & \frac{ \partial l^{(2)}(\thmz)}{\partial \sigma_{u,0}^2} \frac{d^T l^{(2)}(\thmz) }{d \textbf{a}_{m,0}} \\
\frac{ \partial l^{(2)}(\thmz)}{\partial \sigma_{u,0}^2} \frac{d l^{(2)}(\thmz) }{d \textbf{a}_{m,0}} & \frac{d l^{(2)}(\thmz) }{d \textbf{a}_{m,0}} \frac{d^T l^{(2)}(\thmz) }{d \textbf{a}_{m,0}}\\
\end{bmatrix}.
\end{split}
\end{align}
The bound on the spectrum estimation will be taken for $W$ frequency samples:
\begin{equation}
    \boldsymbol{\omega} = [\omega_1,\dots,\omega_W]^T,
\end{equation}
and thus the true vector of interest and the pseudo-true parameter vector are given by:
\begin{equation} \label{eq: spectrum true and pseudo true}
    \boldsymbol{\varphi} = 
    \begin{bmatrix}
\varphi(\omega_1)\\
\dots\\
\varphi(\omega_W)
\end{bmatrix},\indent 
\boldsymbol{\varphi}^{(m)}_0 = 
    \begin{bmatrix}
\varphi_0 ^{(m)}(\omega_1)\\
\dots\\
\varphi_0 ^{(m)}(\omega_W)
\end{bmatrix},
\end{equation}
respectively. The elements of the gradient of the pseudo-true parameter vector $\boldsymbol{\varphi}^{(m)}_0$ w.r.t. $\boldsymbol{\theta}^{(m)}_{0}$, are given by 
(see \cite[Section 5.2.]{porat1994digital}):
\begin{equation} \label{eq: spetrum pseudo-true gradient}\begin{split}
   &\left [\frac{\diff \bvarphi_0^{(m)}}{\diff \thmz}\right ]_{l,1} = \frac{1}{\sigma_{u,0}^2},\\ 
   &\left [\frac{\diff \bvarphi_0^{(m)}}{\diff \thmz} \right ]_{l,k} = -\frac{e^{-j\omega_l(m+2-k)} }{a^{(m)}_0(e^{j\omega_l})} -\frac{e^{j\omega_l(m+2-k)} }{a^{(m)}_0(e^{-j\omega_l})}  ,\\ & l=1,\dots,W,  k=2,\dots,m+1, \end{split}
\end{equation}
where 
\begin{align}
    a^{(m)}_0(e^{j\omega}) = 1+\sum_{k=1}^m [\avec_{m,0}]_{m+1-k} \cdot e^{-j\omega k} 
\end{align}
Finally, substituting (\ref{eq: spectrum MCRB matrices}), (\ref{eq: spectrum true and pseudo true}), and (\ref{eq: spetrum pseudo-true gradient}) in (\ref{eq: Bm 2}) gives the bound for the trace of the MSE for spectrum estimation
\begin{equation} \label{eq: MCRB spectrum}\begin{split}
     B^{(m)}(\boldsymbol{\omega}) &= \left \| \boldsymbol{\varphi} - \boldsymbol{\varphi}^{(m)}_0 \right \|^2 \\
     &+ tr \left( \frac{\diff \bvarphi_0^{(m)}}{\diff \thmz}\textbf{MCRB}\left (\boldsymbol{\theta}^{(m)}_{0} \right) \left (\frac{\diff \bvarphi_0^{(m)}}{\diff \thmz}\right )^T \right).
\end{split}
\end{equation}
It is composed of two terms. The first is the norm squared of the bias between the pseudo-true spectrum of model AR($m$) and the true spectrum. The second term is dependent on the MCRB and bounds the covariance of the estimation. 
\subsubsection{The Cram\'{e}r-Rao bound}
Whittle's formula \cite{porat1994digital} is used to obtain a bound for the mean variance of the log spectrum estimation:
{\small
\begin{equation}\label{eq: CRB}
    \frac{1}{2\pi} \int _{-\pi} ^{\pi} {\rm E } _g \left [ \left(\log \hat{S}(e^{j\omega}) - \log S(e^{j\omega}) \right)^2 \right] d\omega \geq \frac{2}{T} \left(p+q+1 \right).
\end{equation}
}
This bound will be shown later in simulations in comparison to the MCRB for various models.
\subsubsection{Model selection} \label{Sec: Model selection}
As described earlier, the bound in (\ref{eq: MCRB spectrum}) can be used for model selection with the criterion in (\ref{optimal model order}):
\begin{equation} \label{eq: optimal spectrum model}
\begin{split}
    m^* &= \arg \minm_{m \in \mathcal{M}} \Big \{ B^{(m)}(\boldsymbol{\omega}) \Big \} \;. 
\end{split}
\end{equation}
Evaluating the bound in (\ref{eq: MCRB spectrum}) requires knowledge of the true spectrum $\boldsymbol{\varphi}$, the pseudo-true vector $\boldsymbol{\theta}^{(m)}_{0}$, and $\textbf{MCRB}\left (\boldsymbol{\theta}^{(m)}_{0} \right)$. Regarding the pseudo-true vector, it is approximated by the Yule-Walker estimate (with correspondence to (\ref{eq: ML type 2}))
\begin{align} \label{eq: theta YW}
\begin{split}
    \thetaYW & \triangleq \left[\hat{\sigma}_{u,m}^2,\hat{\avec}_{m}^T\right]^T, \quad \hat{\sigma}_{u,m}^2 = \hat{r}_0+\sum_{k=1}^m \hat{r}_k \left[\hat{\avec}_{m}\right]_{m+1-k}\\
    \hat{\avec}_{m}& = - \Pmat_m \hat{\textbf{R}}^{-1}_{m} \hat{\rvec}_m, \quad \hat{\rvec}_m=\left[\hat{r}_1,\dots,\hat{r}_m\right]^T, 
\end{split}
\end{align}
where the permutation matrix $\Pmat_m$ is defined by 
\begin{align}
    \Pmat_m = \begin{bmatrix}
0 & \dots & 0 & 1\\
\dots & 0 & \dots & 0\\
0 & 1 & 0 & \dots\\
1 & 0 & \dots & 0\\
\end{bmatrix} \in \mathbb{R}^{m \times m}.
\end{align}
The true spectrum $\bvarphi$ is approximated by the highest model order spectrum estimation $\bvarphi^{(M)}\left(\hat{\boldsymbol{\theta}}^{(M)}_{YW}\right)$, where its elements are
{\footnotesize
\begin{align} \label{eq: M order spectrum}
    \begin{split}
\left[\bvarphi^{(M)}\left(\hat{\boldsymbol{\theta}}^{(M)}_{YW}\right)\right]_l & = \log \left(\frac{\hat{\sigma}_{u,M}^2}{|1+\sum_{k=1}^M\left[\hat{\avec}_{M}\right]_{M+1-k} e^{-j\omega_l k} |^2}\right),\\ 
        &l=1,\dots,W
    \end{split}
\end{align}
}
and the MCRB by sample estimation of the matrices in (\ref{eq: spectrum MCRB matrices}):
\begin{align} \label{eq: spectrum MCRB matrices est}
\begin{split}
     &\hat{\mathbf{A}}(\thetaYW)  =   \begin{bmatrix}
\frac{ \partial^2 l^{(2)}(\thetaYW)}{\partial \hat{\sigma}_{u,m}^2 \partial \hat{\sigma}_{u,m}^2} & \frac{ d^2 l^{(2)}(\thetaYW)}{d \hat{\avec}^T_{m}\partial \hat{\sigma}_{u,m}^2 } \\
\frac{ d^2 l^{(2)}(\thetaYW)}{d \hat{\avec}_{m}\partial \hat{\sigma}_{u,m}^2 }  & \frac{ d^2 l^{(2)}(\thetaYW)}{d \hat{\avec}_{m} d^T \hat{\avec}_{m}} \\
\end{bmatrix}, \\
&\hat{\mathbf{B}}(\thetaYW) =  
\begin{bmatrix} 
\left(\frac{ \partial l^{(2)}(\thetaYW)}{\partial \hat{\sigma}_{u,m}^2}\right)^2 & \frac{ \partial l^{(2)}(\thetaYW)}{\partial \hat{\sigma}_{u,m}^2} \frac{d^T l^{(2)}(\thetaYW) }{d \hat{\avec}_{m}} \\
\frac{ \partial l^{(2)}(\thetaYW)}{\partial \hat{\sigma}_{u,m}^2} \frac{d l^{(2)}(\thetaYW) }{d \hat{\avec}_{m}} & \frac{d l^{(2)}(\thetaYW) }{d \hat{\avec}_{m}} \frac{d^T l^{(2)}(\thetaYW) }{d \hat{\avec}_{m}}\\
\end{bmatrix},\\
&\Cmat\left({\thetaYW}\right) = \hat{\mathbf{A}}^{-1}(\thetaYW)\hat{\mathbf{B}}(\thetaYW) \hat{\mathbf{A}}^{-1}(\thetaYW).
\end{split}    
\end{align}
The resulted MCRB-based criterion is given by substituting (\ref{eq: theta YW}),(\ref{eq: M order spectrum}) and (\ref{eq: spectrum MCRB matrices est}) in (\ref{eq: B 2 est}):
\begin{align} \label{eq: spectrum B est}
    \begin{split}
        \hat{B}^{(m)} &= \left \| \bvarphi^{(M)}\left(\hat{\boldsymbol{\theta}}^{(M)}_{YW}\right) - \bvarphi^{(m)}\left(\thetaYW\right) \right \|^2 \\
&+ tr \left( \frac{\diff \bvarphi_0^{(m)}}{\diff \thetaYW}\Cmat\left(\thetaYW\right) \left (\frac{\diff \bvarphi_0^{(m)}}{\diff \thetaYW}\right )^T \right),
    \end{split}
\end{align}
and the estimator of $\bvarphi$ based on the selected model order $m^*$ is
{\footnotesize
\begin{align}
\begin{split}
     \hat{\bvarphi}^{(m*)} &= \bvarphi^{(m*)}\left(\hat{\boldsymbol{\theta}}^{(m*)}_{YW}\right),\\
     \left[\bvarphi^{(m*)}\left(\hat{\boldsymbol{\theta}}^{(m*)}_{YW}\right)\right]_l &= \log \left(\frac{\hat{\sigma}_{u,m*}^2}{|1+\sum_{k=1}^{m*}\left[\hat{\avec}_{m*}\right]_{m*+1-k} e^{-j\omega_l k} |^2}\right),\\ 
        &l=1,\dots,W.
\end{split}
\end{align}
}
\section{Simulations}\label{Simulations}

\subsection{DOA Estimation Under Unknown Clutter/Interference}
The MCRB-based criterion in Section \ref{sec: DOA_Estimation} and its MSE performance are analyzed and compared to MDL, AIC, and AICc \cite{Stoica}:
\begin{equation}\label{MDL 1}
    \begin{split}
      MDL(m) = - 2\hat{L}^{(m)} + P_m \log T, 
    \end{split}
    \end{equation}
    \begin{equation}\label{AIC 1}
         AIC(m) = -2 \hat{L}^{(m)} + 2 P_m,
    \end{equation}
    \begin{equation}\label{AICc 1}
        AIC_c(m) =  -2 \hat{L}^{(m)} + \frac{2 T}{(T-1 - P_m)} P_m.
    \end{equation}
Using the first training set to evaluate $\Rxm$, the maximized log-likelihood function at $\Rxm$ (and number of interferences $m$) is \cite{Wax}:
\begin{equation}
    \hat{L}^{(m)} = C + T \sum_{i=m+1}^N \log \left (\frac{\hat{\lambda}_i}{\sm} \right ),
\end{equation}
where $C$ is a constant w.r.t. $\Rxm$. The unknown parameters for this model include the eigenvalues and eigenvectors corresponding to interference and noise subspace, and the noise variance. Thus, the number of unknown parameters is \cite{Wax}:
\begin{equation}
    P_m = m(2N-m) + 1.
\end{equation}

\begin{figure}[htb]
\centering 
\includegraphics[width=0.48\textwidth]{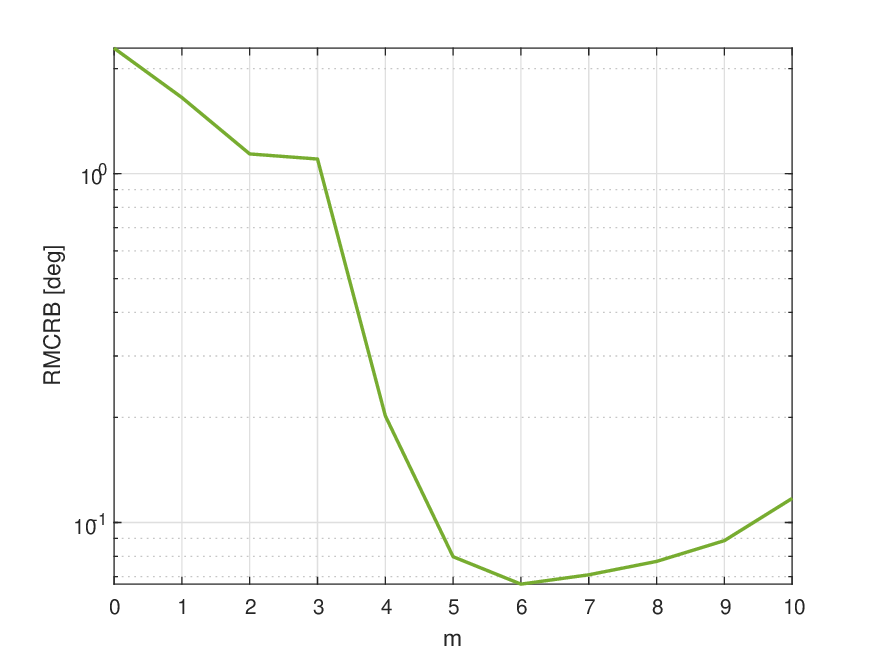}
\caption{Root MCRB, INR=45 dB, SNR=30 dB, $T=12,D=10,$ ULA with $N=11$ sensors.}
\label{fig: 0 MCRB 301023}
\end{figure}

\begin{figure}[htb]
\centering 
\includegraphics[width=0.48\textwidth]{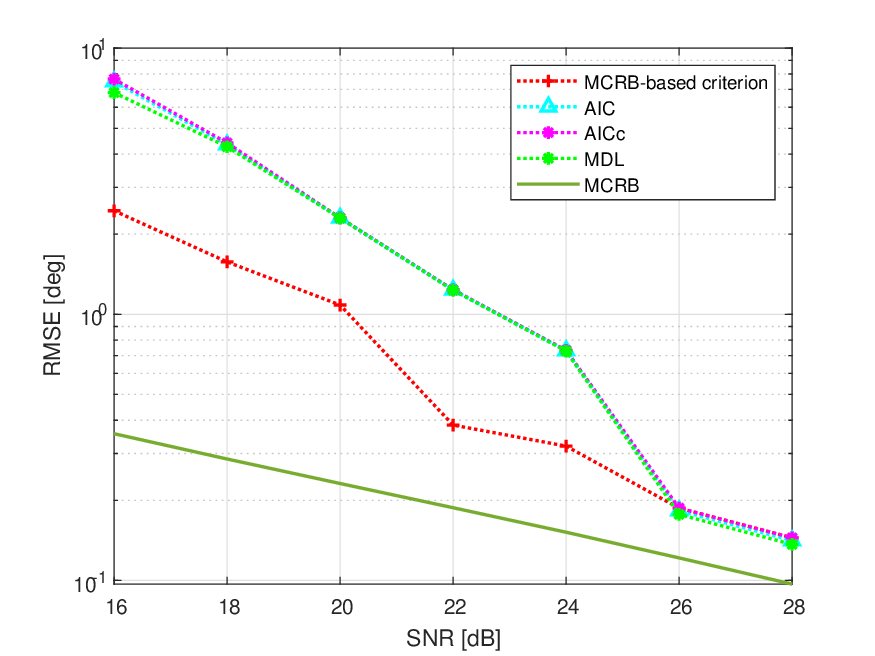}
\caption{DOA estimation versus SNR, INR = SNR+15dB, $T=12,D=10$, ULA with $N=11$ sensors.}
\label{fig: 1 doa 3105}
\end{figure}

\begin{figure}[htb]
\centering 
\includegraphics[width=0.48\textwidth]{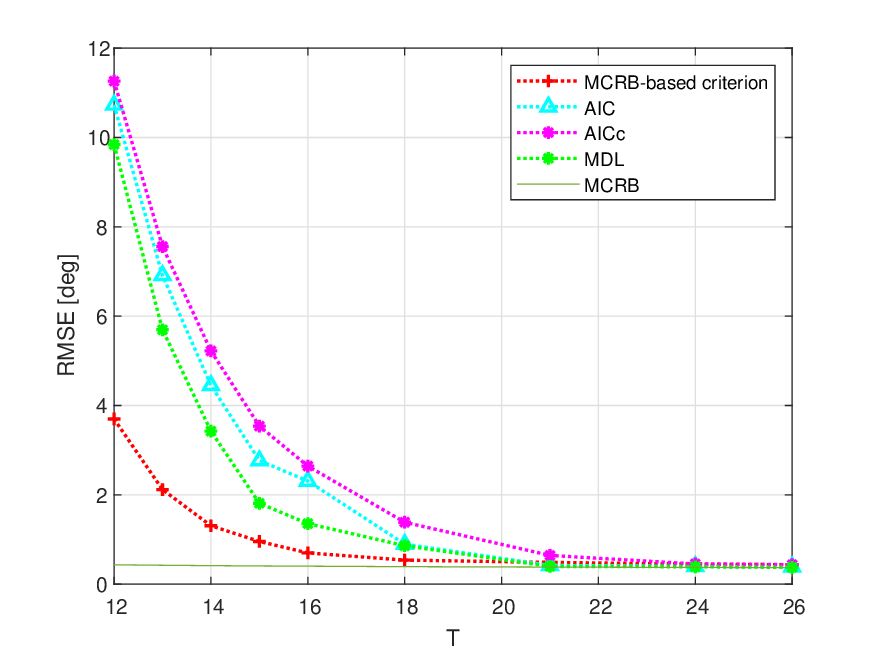}
\caption{DOA estimation versus number of training samples $T$, INR = 30dB, SNR = 15dB, $D=10$, ULA with $N=11$ sensors.}
\label{fig: 2 doa 0609}
\end{figure}

Consider a uniform linear array of $N=11$ sensors with half a wavelength inter-element spacing. The source is considered narrowband and far-field, which
results in steering vector entries \cite{van2004optimum}
\begin{align}
\begin{split}
    [\textbf{a}(\psi)]_n &=\frac{1}{\sqrt{N}}\exp{\left(j \pi \left(n-\frac{N-1}{2}\right) \sin\psi  \right)},\\
    &n=0,\dots,N-1.
\end{split}
\end{align}
The target is at direction $\psi_d=0$, with clutter of $P=6$ interferences, at directions $\boldsymbol{\psi}_c^{(P)} =[-46 ^{\circ},-43^{\circ},-40^{\circ},40^{\circ},43^{\circ},$\newline $46^{\circ}]^T,$ and the interference-to-noise ratio (INR), $INR=\frac{\sigma_c^2}{\sigma^2}$, is greater by 15 dB than signal-to-noise ratio (SNR), $SNR=\frac{|s_d|^2}{\sigma^2}$. 
Fig. \ref{fig: 0 MCRB 301023} displays the MCRB in (\ref{eq: B m DOA under clutter}) versus $m$. For underfitted models $m<P$ some interferences are ignored when estimating $\Rxm$, which results in large misspecified bounds. The MCRB decreases as $m$ increases and more interferences are assumed. The minimal MCRB is at $m=P$, true model order. For overfitted models $m>P$ all interferences are assumed, and the MCRB increases slightly as $m$ increases. Thus, the MCRB, a MSE performance tool, also serves as model selection tool to minimize estimation MSE. 
MSE performance of all criteria are shown in Fig. \ref{fig: 1 doa 3105} for $T=12$, along with the minimal MCRB in (\ref{eq: B m DOA under clutter}) over $m=0,\dots,N-1$. It can be seen that the RMSE's of all the considered criteria are lower bounded by the MCRB. Moreover, the MCRB-based criterion outperforms the others for all SNRs, with SNR threshold lower by 4 dB than the others. 
Estimation of the MCRB using the two measurement sets and minimizing w.r.t. the number of interferences $m$ trade-offs between assuming low and high order clutter. It is focused on the purpose of DOA estimation through the MCRB, while the other criteria only select the number of interferences.
Fig. \ref{fig: 2 doa 0609} presents the RMSE with $SNR=15$dB as a function of the number of training samples $T$. It can be seen that the MCRB-based criterion outperforms the others for small $T$, and as $T$ grows, all criteria performance reach the MCRB, for the estimate covariance matrix converges to the true covariance matrix. It shows that the proposed model selection criterion uses the secondary set better to estimate the DOA, and that effectively less samples are needed in comparison to the other criteria to acquire the same MSE.

\subsection{Auto-Regressive Model Selection}
The criterion in Section \ref{sec: spectrum estimation} for model selection and spectrum estimation will be compared in terms of MSE performance with
MDL, AIC and AICc. For auto-regressive model selection, Section 6.9 in \cite{porat1994digital} evaluates: 
\begin{equation} \label{eq: AIC}
    AIC(m) = T \left (\log\hat{r}_0 + \sum_{n=1}^m\log \left (1-\hat{K}_n^2 \right ) \right ) + 2(m+1),
\end{equation}
\begin{align} \label{eq: MDL}
\begin{split}
     MDL(m) &= T \left (\log\hat{r}_0 + \sum_{n=1}^m\log \left (1-\hat{K}_n^2 \right ) \right )\\
     &+ (m+1) \log T,
\end{split}
\end{align}
and by adjusting the penalty:
\begin{align} \label{eq: AIC_c} \begin{split}
    AIC_c(m) &= T \left (\log\hat{r}_0 + \sum_{n=1}^m\log \left (1-\hat{K}_n^2 \right ) \right )\\ 
    &+ \frac{2(T+1)}{(T - (m+1))} (m+1),
\end{split}
\end{align}
with $\hat{r}_0$ being the zero order biased sample covariance, and $\hat{K}_n$ the $n$th order partial correlation coefficient estimate from the Levinson-Durbin algorithm.

\begin{figure}[htb]
\centering 
\includegraphics[width=0.48\textwidth]{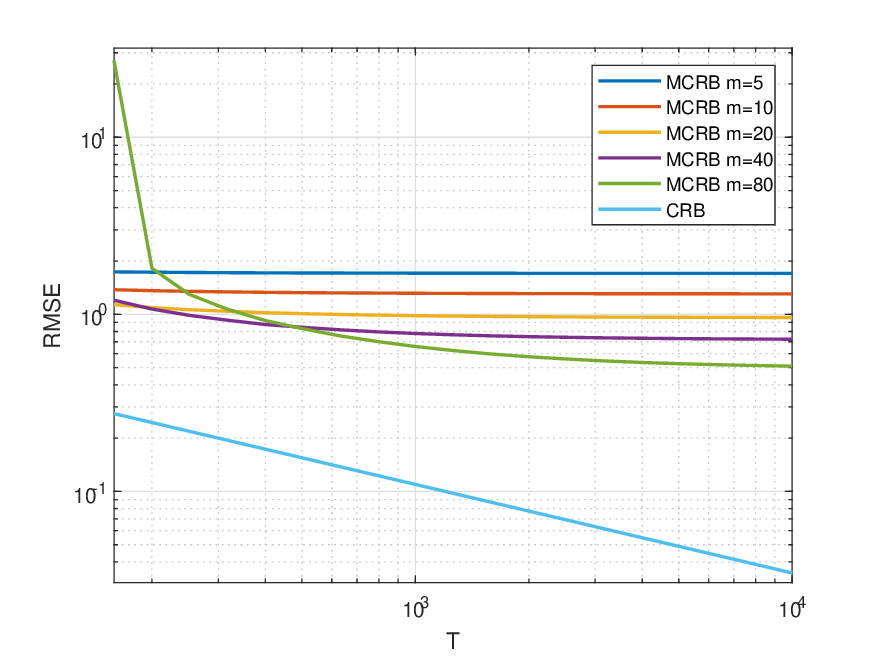}
\caption{MCRB for various model orders, $m=[5,10,20,40,80]$, and CRB, true model MA(5).}
\label{fig:1 spectrum}
\end{figure}

Simulations were taken with a Laplacian MA(5) true model (\ref{eq: Arma}), where
\begin{equation} \label{eq: true model}\begin{split}
       \sigma_u^2=1, \quad 
\textbf{b}_t = \frac{1}{6} \onevec_6
\end{split}
\end{equation}
with models AR($m$) up to order $M=80$. So, the misspecification is in the parameters model and the innovation process. As described earlier, to obtain the pseudo true vector $\boldsymbol{\theta}^{(m)}_{0}$, empirical expectation of the Yule-Walker estimation with large $T$ was used. 
The MCRB were evaluated by empirical expectation over $K=10^3$ trials and $W=1000$ frequencies in $\omega \in [0,\pi]$. Note that the bounds are divided by $W$ to obtain the mean bound of the MSE, for correspondence with (\ref{eq: CRB}). 
Fig. \ref{fig:1 spectrum} shows the MCRB in (\ref{eq: MCRB spectrum}) for model orders $m=[5,10,20,40,80]$ versus the number of samples $T$. As can be seen, the MCRB for the models results in a large bias contribution (independent in $T$), because the models are misspecified and the true spectrum cannot be represented by an AR spectrum estimation. For small $T$s the covariance term is larger than the bias term and dominant, while for large $T$s it is relatively small and thus negligible. For reference, the CRB for the perfectly specified model is smaller than the MCRBs, with zero bias contribution and linear log scaled decreasing covariance. As the model order $m$ enlarges, the model is more complex and the bias contribution decreases, while the covariance term increases. For this case, the optimal model to select in terms of MSE is small $m$s for lower $T$s, and $m=80$ for larger $T$s, as potentially its estimation errors are the lowest.


\begin{figure}[htb]
\centering 
\includegraphics[width=0.48\textwidth]{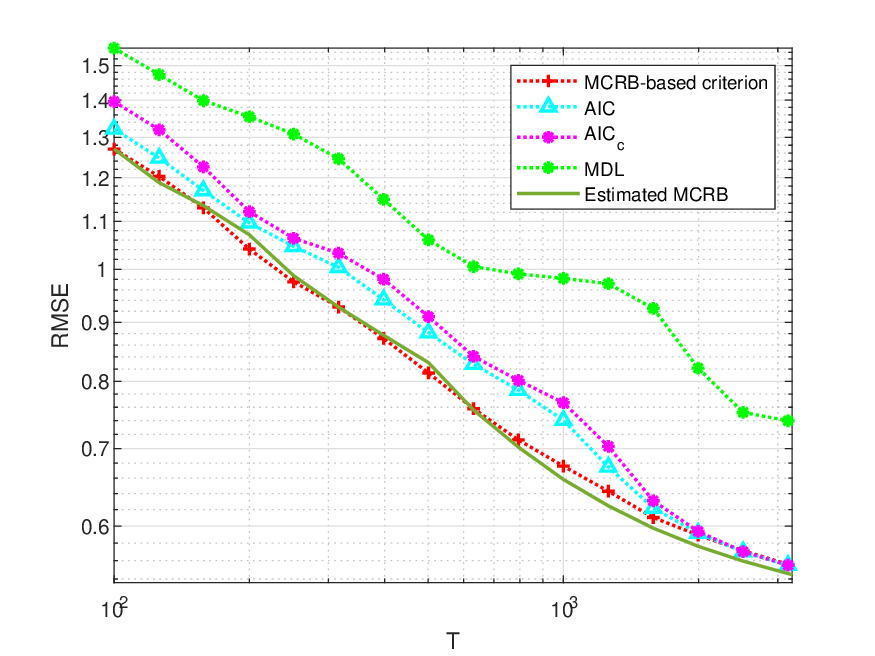}
\caption{RMSE for AR spectrum estimation, $m=[5,10,20,40,80]$, true model MA(5).}
\label{fig:3 spectrum}
\end{figure}

MSE performance of the criteria (\ref{eq: spectrum B est}), (\ref{eq: AIC})-(\ref{eq: AIC_c}) for model selection and estimation are presented in Fig. \ref{fig:3 spectrum}, along with the estimated minimal MCRB. 
As can be seen, the MCRB-based criterion outperforms the other criteria, where its MSE coincides with the minimal MCRB. 
As $T$ enlarges, and AIC and AICc criteria tend to overfit, they converge to the selection of $m=80$. Being the model order with the minimal MCRB, the MSE performance of AIC and AICc converge to the minimal MCRB as well. MDL with the larger penalty term also chooses smaller order models, resulting in larger MSE.



\section{Conclusion}
\label{Conclusion}
In this paper, a new model selection approach for parameter estimation is proposed. In this approach, the MCRB for parameter estimation based on a family of models is computed and minimized for model selection. The MCRB takes into account the effects of both over fitting and model misspecification. The method focuses on better estimation of the parameters of interest, instead of true model selection. It was applied to DOA estimation under unknown clutter/interference and spectrum estimation under unknown ARMA model. In a DOA estimation problem under unknown clutter, the method gained lower threshold SNR in 4 dB against other criteria, and lower RMSE for various number of samples.
In a spectrum estimation problem, the MCRB was derived for misspecified AR($m$) assumed models. It provided understanding of the influence of misspecification in the filter on estimation performance. The MCRB criterion used this analysis to select the model s.t. the spectrum MSE is minimized. It showed to perform better against other criteria for various number of samples, and to its MSE to coincide with the minimal MCRB. 

This work can be continued in several directions. 
With the intensive research of machine learning algorithms, and their applicability to various research fields, the MCRB and the proposed model selection method can give insights about inference selection and its fit to the unknown generative model of the data. Determining of the network architecture or its parameters, such as number of layers, number of parameters, and activation functions, can be considered from a parameter estimation perspective under misspecified modeling. Also, a similar approach for model selection can be suggested for the misspecified Bayesian framework, in which the model parameters are random, using misspecified Bayesian performance bounds.

\appendices

\bibliographystyle{IEEEtran}
\bibliography{refs3}

\begin{thebibliography}{10}
\providecommand{\url}[1]{#1}
\csname url@samestyle\endcsname
\providecommand{\newblock}{\relax}
\providecommand{\bibinfo}[2]{#2}
\providecommand{\BIBentrySTDinterwordspacing}{\spaceskip=0pt\relax}
\providecommand{\BIBentryALTinterwordstretchfactor}{4}
\providecommand{\BIBentryALTinterwordspacing}{\spaceskip=\fontdimen2\font plus
\BIBentryALTinterwordstretchfactor\fontdimen3\font minus
  \fontdimen4\font\relax}
\providecommand{\BIBforeignlanguage}[2]{{%
\expandafter\ifx\csname l@#1\endcsname\relax
\typeout{** WARNING: IEEEtran.bst: No hyphenation pattern has been}%
\typeout{** loaded for the language `#1'. Using the pattern for}%
\typeout{** the default language instead.}%
\else
\language=\csname l@#1\endcsname
\fi
#2}}
\providecommand{\BIBdecl}{\relax}
\BIBdecl

\bibitem{Ding}
J.~Ding, V.~Tarokh, and Y.~Yang, ``Model selection techniques—an overview,''
  \emph{IEEE Signal Process. Mag.}, vol. 35, no 6, pp. 16--34, 2018.

\bibitem{Stoica}
P.~Stoica and Y.~Selen, ``Model-order selection: A review of information
  criterion rules,'' \emph{IEEE Signal Process. Mag.}, vol. 21, no 4, pp.
  36--47, 2004.

\bibitem{Kadane}
J.~B. Kadane and N.~A. Lazar, ``Methods and criteria for model selection,''
  \emph{J. Amer. Statist. Assoc.}, vol. 99, no. 465, pp. 279--290, 2004.

\bibitem{AIC}
H.~Akaike, ``A new look at the statistical model identification,'' \emph{IEEE
  Trans. Autom. Control}, vol.~19, p. 716–723, Dec. 1974.

\bibitem{BIC}
G.~Schwarz, ``Estimating the dimension of a model,'' \emph{Ann. Statist.},
  vol.~6, p. 461–464, Mar. 1978.

\bibitem{MDL1}
J.~Rissanen, ``Modeling by shortest data description,'' \emph{Automatica},
  vol.~14, p. 465–471, Sep. 1978.

\bibitem{MDL2}
------, ``Estimation of structure by minimum description length,''
  \emph{Circuits Syst. Signal Process.}, vol.~1, p. 395–406, Sep. 1982.

\bibitem{MDL3}
A.~Barron, J.~Rissanen, and B.~Yu, ``The minimum description length principle
  in coding and modeling,'' \emph{IEEE Trans. Inf. Theory}, vol.~44, p.
  2743–2760, Oct. 1998.

\bibitem{MDL4}
M.~H. Hansen and B.~Yu, ``Model selection and the principle of minimum
  description length,'' \emph{J. Amer. Statist. Assoc.}, vol.~96, pp. 746--774,
  Jun. 2001.

\bibitem{Rosenthal}
N.~E. Rosenthal and J.~Tabrikian, ``Model selection via misspecified
  {C}ramér-{R}ao bound minimization,'' in \emph{ICASSP 2022 - 2022 IEEE
  International Conference on Acoustics, Speech and Signal Processing
  (ICASSP)}, pp. 5762--5766.

\bibitem{Stefano3}
S.~Fortunati, F.~Gini, M.~Greco, and C.~Richmond, ``Performance bounds for
  parameter estimation under misspecified models fundamental findings and
  applications,'' \emph{IEEE Signal Process. Mag.}, vol. 34, no. 6, pp.
  142--157, 2017.

\bibitem{CRB1}
H.~{C}ram\'{e}r, \emph{Mathematical Methods of Statistics.}\hskip 1em plus
  0.5em minus 0.4em\relax Princeton, NJ, USA, Princeton Univ. Press, 1946.

\bibitem{CRB2}
C.~R. Rao, ``Information and accuracy attainable in the estimation of
  statistical parameters,'' \emph{Bull. Calcutta Math. Soc.}, vol.~37, pp.
  81--91, 1945.

\bibitem{White}
H.~White, ``Maximum likelihood estimation of misspecified models,''
  \emph{Econometrica}, vol.~50, pp. 1--25, 1982.

\bibitem{White_Book}
------, \emph{Estimation, Inference and Specification Analysis.}\hskip 1em plus
  0.5em minus 0.4em\relax Cambridge University Press, 1996.

\bibitem{Noam_Tabrikian}
Y.~Noam and J.~Tabrikian, ``Marginal likelihood for estimation and detection
  theory,'' \emph{IEEE Trans. Signal Process.}, vol.~55, no.~8, pp. 3963--3974,
  2007.

\bibitem{Vuong86}
Q.~H. Vuong, ``{C}ram$\acute{\text{e}}$r–{R}ao bounds for misspecified
  models,'' \emph{Div. of the Humanities and Social Sci., California Inst. of
  Technol., Pasadena, CA, USA}, 1986.

\bibitem{Richmond}
C.~D. Richmond and L.~L. Horowitz, ``Parameter bounds on estimation accuracy
  under model misspecification,'' \emph{IEEE Trans. Signal Process.}, vol. 63,
  no. 9, pp. 2263--2278, 2015.

\bibitem{Richmond2}
------, ``Parameter bounds under misspecified models,'' \emph{Proc. Conf.
  Signals, Systems and Computers (Asilomar)}, pp. 176--180, 2013.

\bibitem{Stefano}
S.~Fortunati, F.~Gini, and M.~Greco, ``The misspecified
  {C}ram$\acute{\text{e}}$r-{R}ao bound and its application to scatter matrix
  estimation in complex elliptically symmetric distributions,'' \emph{IEEE
  Trans. Signal Process.}, vol. 64, no. 9, p. 2387–2399, 2016.

\bibitem{Wang_Gini}
P.~Wang, T.~Koike-Akino, M.~Pajovic, P.~V. Orlik, W.~Tsujita, and F.~Gini,
  ``Misspecified {CRB} on parameter estimation for a coupled mixture of
  polynomial phase and sinusoidal {FM} signals,'' in \emph{2019 IEEE
  International Conference on Acoustics, Speech and Signal Processing
  (ICASSP)}, pp. 5302--5306.

\bibitem{9537597}
L.~T. Thanh, K.~Abed-Meraim, and N.~L. Trung, ``Misspecified
  {C}ram$\acute{\text{e}}$r–{R}ao bounds for blind channel estimation under
  channel order misspecification,'' \emph{IEEE Trans. Signal Process.},
  vol.~69, pp. 5372--5385, 2021.

\bibitem{MCPHEE2023108872}
\BIBentryALTinterwordspacing
H.~McPhee, L.~Ortega, J.~Vilà-Valls, and E.~Chaumette, ``On the accuracy
  limits of misspecified delay-{D}oppler estimation,'' \emph{Signal
  Processing}, vol. 205, p. 108872, 2023. [Online]. Available:
  \url{https://www.sciencedirect.com/science/article/pii/S016516842200411X}
\BIBentrySTDinterwordspacing

\bibitem{https://doi.org/10.48550/arxiv.2203.03398}
\BIBentryALTinterwordspacing
M.~Hellkvist, A.~Özçelikkale, and A.~Ahlén, ``Estimation under model
  misspecification with fake features,'' 2022. [Online]. Available:
  \url{https://arxiv.org/abs/2203.03398}
\BIBentrySTDinterwordspacing

\bibitem{Stefano2}
S.~Fortunati, F.~Gini, and M.~Greco, ``The constrained misspecified
  {C}ram$\acute{\text{e}}$r-{R}ao bound,'' \emph{IEEE Signal Process. Lett.},
  vol. 23, no. 5, p. 718–721, 2016.

\bibitem{Fortunati}
S.~Fortunati, ``Misspecified {C}ram$\acute{\text{e}}$r–{R}ao bounds for
  complex unconstrained and constrained parameters,'' \emph{Proc. Eur. Signal
  Process. Conf. (EUSIPCO)}, 2017.

\bibitem{Huber}
P.~J. Huber, ``The behavior of maximum likelihood estimates under nonstandard
  conditions,'' \emph{Proc. 5th Berkley Symp. Math. Statist. Probab}, p.
  221–233, 1967.

\bibitem{Kelly}
E.~Kelly, ``An adaptive detection algorithm,'' \emph{IEEE Transactions on
  Aerospace and Electronic Systems}, vol. AES-22, no.~2, pp. 115--127, 1986.

\bibitem{Maio}
E.~Conte, A.~De~Maio, and G.~Ricci, ``Glrt-based adaptive detection algorithms
  for range-spread targets,'' \emph{IEEE Trans. Signal Process.}, vol.~49,
  no.~7, pp. 1336--1348, 2001.

\bibitem{van2004optimum}
H.~L. Van~Trees, \emph{Optimum Array Processing: Part IV of Detection,
  Estimation, and Modulation Theory}.\hskip 1em plus 0.5em minus 0.4em\relax
  John Wiley \& Sons, 2004.

\bibitem{porat1994digital}
B.~Porat, ``Digital processing of random signals: theory and methods,'' 1994.

\bibitem{Wax}
M.~Wax and T.~Kailath, ``Detection of signals by information theoretic
  criteria,'' \emph{IEEE Trans. Acoust., Speech, Signal Process.}, vol.~33,
  no.~2, pp. 387--392, 1985.

\end{thebibliography}

\end{document}